\documentclass[twocolumn,aps,superscriptaddress,showpacs]{revtex4}
\usepackage{amsmath,bm}
\usepackage{graphicx}

\begin{document}

\title{Effects of isospin and momentum dependent interactions on thermal
properties of asymmetric nuclear matter}
\author{Jun Xu}
\affiliation{Institute of Theoretical Physics, Shanghai Jiao Tong University, Shanghai
200240, China}
\author{Lie-Wen Chen}
\affiliation{Institute of Theoretical Physics, Shanghai Jiao Tong University, Shanghai
200240, China}
\affiliation{Center of Theoretical Nuclear Physics, National Laboratory of Heavy-Ion
Accelerator, Lanzhou, 730000, China}
\author{Bao-An Li}
\affiliation{Department of Physics, Texas A\&M University-Commerce, Commerce, TX
75429-3011, USA}
\author{Hong-Ru Ma}
\affiliation{Institute of Theoretical Physics, Shanghai Jiao Tong University, Shanghai
200240, China}

\begin{abstract}
Thermal properties of asymmetric nuclear matter are studied within
a self-consistent thermal model using an isospin and momentum
dependent interaction (MDI) constrained by the isospin diffusion
data in heavy-ion collisions, a momentum-independent interaction
(MID), and an isoscalar momentum-dependent interaction (eMDYI). In
particular, we study the temperature dependence of the
isospin-dependent bulk and single-particle properties, the
mechanical and chemical instabilities, and liquid-gas phase
transition in hot asymmetric nuclear matter. Our results indicate
that the temperature dependence of the equation of state and the
symmetry energy are not so sensitive to the momentum dependence of
the interaction. The symmetry energy at fixed density is found to
generally decrease with temperature and for the MDI interaction
the decrement is essentially due to the potential part. It is
further shown that only the low momentum part of the
single-particle potential and the nucleon effective mass increases
significantly with temperature for the momentum-dependent
interactions. For the MDI interaction, the low momentum part of
the symmetry potential is significantly reduced with increasing
temperature. For the mechanical and chemical instabilities as well
as the liquid-gas phase transition in hot asymmetric nuclear
matter, our results indicate that the boundary of these
instabilities and the phase-coexistence region generally shrink
with increasing temperature and is sensitive to the density
dependence of the symmetry energy and the isospin and momentum
dependence of the nuclear interaction, especially at higher
temperatures.
\end{abstract}

\maketitle

\section{Introduction}

The recent progress in developing advanced radioactive beam
facilities offers a great opportunity to explore in terrestrial
laboratories properties of nuclear matter and/or nuclei with large
isospin asymmetries. As a result, studies on the role of the
isospin degree of freedom have recently attracted much attention
in both nuclear physics and astrophysics. The ultimate goal of
such studies is to extract information on the isospin dependence
of in-medium nuclear effective interactions as well as the
equation of state (EOS) of isospin asymmetric nuclear matter. The
latter, especially the nuclear symmetry energy term, is important
for understanding not only many aspects of nuclear physics, but
also a number of important issues in astrophysics
\cite{LiBA98,LiBA01b,Dan02a,Lat00,Lat01,Lat04,Bar05,Ste05a}.
Information about the symmetry energy at zero temperature is
important for determining ground state properties of exotic nuclei
and properties of cold neutron stars at $\beta $-equilibrium,
while the symmetry energy or symmetry free energy of hot
neutron-rich matter is important for understanding the liquid-gas
(LG) phase transition of asymmetric nuclear matter, the dynamical
evolution of massive stars and the supernova explosion mechanisms.

Although the nuclear symmetry energy at normal nuclear matter density for
cold asymmetric nuclear matter is known to be around $30$ MeV from the
empirical liquid-drop mass formula \cite{Mey66,Pom03}, its values at other
densities, especially at supra-normal densities, are poorly known \cite%
{LiBA98,LiBA01b}. Predictions based on various many-body theories
differ widely at both low and high densities \cite{Bom01,Die03}.
Empirically, the incompressibility of asymmetric nuclear matter is
essentially undetermined \cite{Shl93}, even though the
incompressibility of symmetric nuclear matter at its saturation
density $\rho _{0}\approx 0.16$ fm$^{-3}$ has been determined to
be $231\pm 5 $ MeV from nuclear giant monopole resonances (GMR)
\cite{You99} and the EOS at densities of $2\rho _{0}<\rho <5\rho
_{0}$ has also been constrained by measurements of collective
flows in nucleus-nucleus collisions \cite{Dan02a}. Fortunately,
heavy-ion reactions, especially those induced by radioactive
beams, provide a unique means to investigate the isospin-dependent
properties of asymmetric nuclear matter, particularly the density
dependence of the nuclear symmetry energy. Indeed, significant
progress has recently been made both experimentally and
theoretically in extracting the information on the behaviors of
nuclear symmetry energy at sub-saturation density from the isospin
diffusion data in heavy-ion collisions from the NSCL/MSU
\cite{Tsa04,Che05a,LiBA05c}. Using the isospin and
momentum-dependent IBUU04 transport model with in-medium
nucleon-nucleon (NN) cross sections, the isospin diffusion data
were found to be consistent with a density-dependent symmetry
energy of $E_{\mathrm{sym}}(\rho )\approx 31.6(\rho /\rho
_{0})^{\gamma }$ with $\gamma =0.69-1.05$ at subnormal density
\cite{Che05a,LiBA05c}, which has led to the extraction of a value
of $L=88\pm 25$ MeV for the slope parameter of the nuclear
symmetry energy at saturation density and a value of
$K_{\mathrm{asy}}=-500\pm 50$ MeV for the isospin-dependent part
of the isobaric incompressibility of isospin asymmetric nuclear
matter \cite{Che05a,LiBA05c,Che05b}. The extracted
symmetry energy further agrees with the symmetry energy $E_{\mathrm{sym}%
}(\rho )=31.6(\rho /\rho _{0})^{0.69}$ recently obtained from the isoscaling
analyses of isotope ratios in intermediate energy heavy ion collisions \cite%
{She07}, which gives $L\approx 65$ MeV and $K_{\mathrm{asy}}\approx -453$
MeV. The extracted value of $K_{\mathrm{asy}}=-500\pm 50$ MeV from the
isospin diffusion data is also consistent with the value $K_{\mathrm{asy}%
}=-550\pm 100$ MeV obtained from recently measured isotopic dependence of
the GMR in even-A Sn isotopes \cite{Gar07}. These empirically extracted
values for $L$ and $K_{\mathrm{asy}}$ represent the best and most stringent
phenomenological constraints available so far on the nuclear symmetry energy
in cold asymmetric nuclear matte at sub-normal densities.

Theoretically, a lot of efforts have been devoted to the study on the
properties of cold asymmetric nuclear matter. However, the properties of hot
asymmetric nuclear matter, especially the temperature-dependence of the
nuclear symmetry energy, has received so far much less attention\cite%
{Che01b,Zuo03,LiBA06c,Xu07,Xu07b,Mou07}. For finite nuclei at temperatures
below about $3$ MeV, the shell structure and pairing as well as vibrations
of nuclear surfaces are important and the symmetry energy was predicted to
increase slightly with the increasing temperature \cite{Don94,Dea95,Dea02}.
Interestingly, an increase by only about $8\%$ in the symmetry energy in the
range of $T$ from $0$ to $1$ MeV was found to affect appreciably the physics
of stellar collapse, especially the neutralization processes \cite{Don94}.
At higher temperatures, one expects the symmetry energy to decrease as the
Pauli blocking becomes less important when the nucleon Fermi surfaces become
more diffused at increasingly higher temperatures \cite%
{Che01b,Zuo03,LiBA06c,Xu07,Xu07b,Mou07}. On the other hand, due to
the van der Waals behavior of the nucleon-nucleon interaction, it
is expected that the so-called LG phase transition should occur in
nuclear matter. Since the early work, see, e.g., Refs.
\cite{Lam78,Fin82,Ber83,Jaq83}, many investigations have been
carried out to explore properties of the nuclear LG phase
transition both experimentally and theoretically over the
last three decades. For a recent review, see, e.g., Refs. \cite%
{Cho04,Das05,Cho06}. Most of these studies focused on investigating features
of the LG phase transition in symmetric nuclear matter. New features of the
LG phase transition in asymmetric nuclear matter are expected. In
particular, in a two-component asymmetric nucleonic matter, there are two
conserved charges of baryon number and the third component of isospin. The
LG phase transition was suggested to be of second order\cite{Mul95}. This
suggestion together with the need to understand better the properties of
asymmetric nuclear matter have stimulated a lot of work recently, see, e.g.,
Refs. \cite%
{LiBA97b,Wan00,Su00,Lee01,Mek05,LiBA01c,Nat02,LiBA02b,Cho03,Sil04,Duc06,Duc07,Xu07b}%
.

While significant progress has been made recently, many
interesting questions about properties of hot asymmetric nuclear
matter remain open. Some of these questions can be traced back to
our poor understanding about the isovector nuclear interaction and
the density dependence of the nuclear symmetry energy
\cite{LiBA98,LiBA01b,Cho06}. With the recent progress on the
constraints of the density dependence of nuclear symmetry energy,
it is therefore interesting to investigate how the constrained
symmetry energy may allow us to better understand the thermal
properties of asymmetric nuclear matter. Moreover, both the
isovector (i.e., the nuclear symmetry potential) and isoscalar
parts of the single nucleon potential should be momentum dependent
due to the non-locality of nucleon-nucleon interaction and the
Pauli exchange effects in many-fermion systems. However, effects
of the momentum-dependent interactions on the thermal properties
of asymmetric nuclear matter have received so far little
theoretical attention \cite{Xu07,Xu07b,Mou07}.

In the present work, we study systematically the effects of isospin and
momentum dependent interactions on the thermal properties of asymmetric
nuclear matter, including the isospin-dependent bulk and single-particle
properties, mechanical and chemical instability, and LG phase transition,
within a self-consistent thermal model using three different interactions.
The first one is the isospin and momentum dependent MDI interaction
constrained by the isospin diffusion data in heavy-ion collisions. The
second one is a momentum-independent interaction (MID) which leads to a
fully momentum independent single nucleon potential, and the third one is an
isoscalar momentum-dependent interaction (eMDYI) in which the isoscalar part
of the single nucleon potential is momentum dependent but the isovector part
of the single nucleon potential is momentum independent by construction. We
note that the MDI interaction is realistic, while the MID and eMDYI
interactions are only used as references in order to explore effects of the
isospin and momentum dependence of the nuclear interactions.

The paper is organized as follows. In Section \ref{model}, we briefly
introduce the MDI, MID and eMDYI interactions and discuss some relevant
thermodynamic quantities. Results on thermal effects on the
isospin-dependent bulk and single-particle properties of asymmetric nuclear
matter, such as the nuclear symmetry energy, the nuclear symmetry potential
and isospin-splitting of nucleon effective mass, are presented in Section %
\ref{bulk}. The mechanical and chemical instabilities of hot neutron-rich
nuclear matter are then discussed in Section \ref{instability}. In Section %
\ref{transition}, we present the results on the LG phase
transition of hot asymmetric nuclear matter. All the results are
calculated from the three interactions, and the effects of isospin
and momentum dependence the nuclear interactions are analyzed.
Finally, a summary is given in Section \ref{summary}.

\section{Theoretical models}

\label{model}

\subsection{Isospin and momentum-dependent MDI interaction}

The isospin- and momentum-dependent MDI interaction is based on a modified
finite-range Gogny effective interaction~\cite{Das03}. In the MDI
interaction, the potential energy density $V(\rho ,T,\delta )$ of a thermal
equilibrium asymmetric nuclear matter at total density $\rho $, temperature $%
T$ and isospin asymmetry $\delta $ is expressed as follows~\cite%
{Das03,Che05a},
\begin{eqnarray}
V(\rho ,T,\delta ) &=&\frac{A_{u}\rho _{n}\rho _{p}}{\rho _{0}}+\frac{A_{l}}{%
2\rho _{0}}(\rho _{n}^{2}+\rho _{p}^{2})+\frac{B}{\sigma +1}\frac{\rho
^{\sigma +1}}{\rho _{0}^{\sigma }}  \notag \\
&\times &(1-x\delta ^{2})+\frac{1}{\rho _{0}}\sum_{\tau ,\tau ^{\prime
}}C_{\tau ,\tau ^{\prime }}  \notag \\
&\times &\int \int d^{3}pd^{3}p^{\prime }\frac{f_{\tau }(\vec{r},\vec{p}%
)f_{\tau ^{\prime }}(\vec{r},\vec{p}^{\prime })}{1+(\vec{p}-\vec{p}^{\prime
})^{2}/\Lambda ^{2}}.  \label{MDIV}
\end{eqnarray}%
In the mean field approximation, Eq. (\ref{MDIV}) leads to the following
single particle potential for a nucleon with momentum $\vec{p}$ and isospin $%
\tau $ in the thermal equilibrium asymmetric nuclear matter, i.e.,~\cite%
{Das03,Che05a}

\begin{eqnarray}
U(\rho ,T,\delta ,\vec{p},\tau ) &=&A_{u}(x)\frac{\rho _{-\tau }}{\rho _{0}}%
+A_{l}(x)\frac{\rho _{\tau }}{\rho _{0}}  \notag \\
&+&B(\frac{\rho }{\rho _{0}})^{\sigma }(1-x\delta ^{2})-8\tau x\frac{B}{%
\sigma +1}\frac{\rho ^{\sigma -1}}{\rho _{0}^{\sigma }}\delta \rho _{-\tau }
\notag \\
&+&\frac{2C_{\tau ,\tau }}{\rho _{0}}\int d^{3}p^{\prime }\frac{f_{\tau }(%
\vec{r},\vec{p}^{\prime })}{1+(\vec{p}-\vec{p}^{\prime })^{2}/\Lambda ^{2}}
\notag \\
&+&\frac{2C_{\tau ,-\tau }}{\rho _{0}}\int d^{3}p^{\prime }\frac{f_{-\tau }(%
\vec{r},\vec{p}^{\prime })}{1+(\vec{p}-\vec{p}^{\prime })^{2}/\Lambda ^{2}}.
\label{MDIU}
\end{eqnarray}%
In the above $\tau =1/2$ ($-1/2$) for neutrons (protons); $\sigma =4/3$; $%
f_{\tau }(\vec{r},\vec{p})$ is the phase space distribution function at
coordinate $\vec{r}$ and momentum $\vec{p}$. The parameters $%
A_{u}(x),A_{l}(x),B,C_{\tau ,\tau },C_{\tau ,-\tau }$ and $\Lambda $ have
been assumed to be temperature independent and are obtained by fitting the
momentum-dependence of $U(\rho ,T=0,\delta ,\vec{p},\tau )$ to that
predicted by the Gogny Hartree-Fock and/or the Brueckner-Hartree-Fock
calculations, the zero temperature saturation properties of symmetric
nuclear matter and the symmetry energy of $31.6$ MeV at normal nuclear
matter density $\rho _{0}=0.16$ fm$^{-3}$ \cite{Das03}. The
incompressibility $K_{0}$ of cold symmetric nuclear matter at saturation
density $\rho _{0}$ is set to be $211$ MeV. The parameters $A_{u}(x)$ and $%
A_{l}(x)$ depend on the $x$ parameter according to
\begin{equation}
A_{u}(x)=-95.98-x\frac{2B}{\sigma +1},~A_{l}(x)=-120.57+x\frac{2B}{\sigma +1}%
.
\end{equation}%
The different $x$ values in the MDI interaction are introduced to vary the
density dependence of the nuclear symmetry energy while keeping other
properties of the nuclear equation of state fixed \cite{Che05a} and they can
be adjusted to mimic predictions on the density dependence of nuclear matter
symmetry energy by microscopic and/or phenomenological many-body theories.
The last two terms of Eq. (\ref{MDIU}) contain the momentum-dependence of
the single-particle potential. The momentum dependence of the symmetry
potential stems from the different interaction strength parameters $C_{\tau
,-\tau }$ and $C_{\tau ,\tau }$ for a nucleon of isospin $\tau $
interacting, respectively, with unlike and like nucleons in the background
fields. More specifically, we use $C_{\tau ,-\tau }=-103.4$ MeV and $C_{\tau
,\tau }=-11.7$ MeV. We note that the MDI interaction has been extensively
used in the transport model for studying isospin effects in intermediate
energy heavy-ion collisions induced by neutron-rich nuclei \cite%
{LiBA04a,Che04,Che05a,LiBA05a,LiBA05b,LiBA06b,Yon06a,Yon06b,Yon07}. In
particular, the isospin diffusion data from NSCL/MSU have constrained the
value of $x$ to be between $0$ and $-1$ for nuclear matter densities less
than about $1.2\rho _{0}$ \cite{Che05a,LiBA05c}, we will thus in the present
work consider the two values of $x=0$ and $x=-1$. We note that the
zero-temperature symmetry energy for the MDI interaction with $x=0$ and $-1$
can be parameterized, respectively, as $31.6(\rho /\rho _{0})^{0.69}$ MeV
and $31.6(\rho /\rho _{0})^{1.05}$ MeV \cite{Che05a}, and thus $x=0$ gives a
softer symmetry energy while $x=-1$ gives a stiffer symmetry energy.

\subsection{Momentum-independent MID interaction}

In the momentum-independent MID interaction, the potential energy density $%
V_{\text{MID}}(\rho ,\delta )$ of a thermally equilibrated asymmetric
nuclear matter at total density $\rho $ and isospin asymmetry $\delta $ is
written as
\begin{equation}
V_{\text{MID}}(\rho ,\delta )=\frac{\alpha }{2}\frac{\rho ^{2}}{\rho _{0}}+%
\frac{\beta }{1+\gamma }\frac{\rho ^{1+\gamma }}{{\rho _{0}}^{\gamma }}+{%
\rho }E_{sym}^{pot}(\rho ,x){\delta }^{2}.  \label{MIDV}
\end{equation}%
The parameters $\alpha $, $\beta $ and $\gamma $ are determined by the
incompressibility $K_{0}$ of cold symmetric nuclear matter at saturation
density $\rho _{0}$ \cite{LiBA97b}
\begin{eqnarray}
\alpha &=&-29.81-46.90\frac{K_{0}+44.73}{K_{0}-166.32}~\text{(MeV)} \\
\beta &=&23.45\frac{K_{0}+255.78}{K_{0}-166.32}~\text{(MeV)} \\
\gamma &=&\frac{K_{0}+44.73}{211.05}
\end{eqnarray}%
and $K_{0}$ is again set to be $211$ MeV as in the MDI interaction. To fit
the MDI interaction at zero temperature, the potential part of the symmetry
energy $E_{sym}^{pot}(\rho ,x)$ is parameterized by \cite{Che05a}
\begin{equation}
E_{sym}^{pot}(\rho ,x)=F(x)\frac{\rho }{\rho _{0}}+\left[ 18.6-F(x)\right] (%
\frac{\rho }{\rho _{0}})^{G(x)}  \label{epotsym}
\end{equation}%
with $F(x=0)=129.981$ MeV, $G(x=0)=1.059$, $F(x=-1)=3.673$ MeV,
and $G(x=-1)=1.569$. We note that the MID interaction reproduce
very well the EOS of isospin-asymmetric nuclear matter with the
MDI interaction at zero temperature for both $x=0$ and $x=-1$. The
single nucleon potential in the MID interaction can be directly
obtained as
\begin{equation}
U_{\text{MID}}(\rho ,\delta ,\tau )=\alpha \frac{\rho }{\rho _{0}}+\beta (%
\frac{\rho }{\rho _{0}})^{\gamma }+U^{\text{asy}}(\rho ,\delta ,\tau ),
\end{equation}%
with
\begin{eqnarray}
U^{\text{asy}}(\rho ,\delta ,\tau ) &=&\left[ 4F(x)\frac{\rho }{\rho _{0}}%
+4(18.6-F(x))(\frac{\rho }{\rho _{0}})^{G(x)}\right] {\tau }{\delta }  \notag
\\
&+&(18.6-F(x))(G(x)-1)(\frac{\rho }{\rho _{0}})^{G(x)}{\delta }^{2}.
\label{Uasy}
\end{eqnarray}%
Therefore, the single nucleon potential in the MID interaction is fully
momentum-independent. It also leads to the fact that the potential energy
density and the single nucleon potential in the MID interaction are
independent of the temperature.

\subsection{Extended MDYI (eMDYI) interaction}

The momentum-dependent part in the MDI interaction is also isospin dependent
while the MID interaction is fully momentum independent. In order to see the
effect of the momentum dependence of the isovector part of the single
nucleon potential (nuclear symmetry potential), we can construct an
isoscalar momentum-dependent interaction, called extended MDYI (eMDYI)
interaction since it has the same functional form as the well-known MDYI
interaction for symmetric nuclear matter \cite{Gal90}. In the eMDYI
interaction, the potential energy density $V_{\text{eMDYI}}(\rho ,T,\delta )$
of a thermally equilibrated asymmetric nuclear matter at total density $\rho
$, temperature $T$ and isospin asymmetry $\delta $ is expressed as
\begin{eqnarray}
V_{\text{eMDYI}}(\rho ,T,\delta ) &=&\frac{A}{2}\frac{\rho ^{2}}{\rho _{0}}+%
\frac{B}{1+\sigma }\frac{\rho ^{1+\sigma }}{{\rho _{0}}^{\sigma }}  \notag \\
&+&\frac{C}{\rho _{0}}\int \int d^{3}pd^{3}p^{\prime }\frac{f_{0}(\vec{r},%
\vec{p})f_{0}(\vec{r},\vec{p}^{\prime })}{1+(\vec{p}-\vec{p}^{\prime
})^{2}/\Lambda ^{2}}  \notag \\
&+&{\rho }E_{sym}^{pot}(\rho ,x){\delta }^{2}.  \label{MDYIV}
\end{eqnarray}%
Here $f_{0}(\vec{r},\vec{p})$ is the phase space distribution function of
\emph{symmetric nuclear matter} at total density $\rho $ and temperature $T$%
. $E_{sym}^{pot}(\rho ,x)$ has the same expression as Eq.~(\ref{epotsym}).
We set $A=\frac{A_{u}+A_{l}}{2}$ and $C=\frac{C_{\tau ,-\tau }+C_{\tau ,\tau
}}{2}$, and $B$, $\sigma $ and $\Lambda $ have the same values as in the MDI
interaction, so that the eMDYI interaction gives the same EOS of asymmetric
nuclear matter as the MDI interaction at zero temperature for both $x=0$ and
$x=-1$. The single nucleon potential in the eMDYI interaction can be
obtained as%
\begin{equation}
U_{\text{eMDYI}}(\rho ,T,\delta ,\vec{p},\tau )=U^{0}(\rho ,T,\vec{p}%
)+U^{asy}(\rho ,\delta ,\tau ),
\end{equation}%
where
\begin{eqnarray}
U^{0}(\rho ,T,\vec{p}) &=&A\frac{\rho }{\rho _{0}}+B(\frac{\rho }{\rho _{0}}%
)^{\sigma }  \notag \\
&+&\frac{2C}{\rho _{0}}\int d^{3}p^{\prime }\frac{f_{0}(\vec{r},\vec{p})}{1+(%
\vec{p}-\vec{p}^{\prime })^{2}/\Lambda ^{2}}
\end{eqnarray}%
and $U^{\text{asy}}(\rho ,\delta ,\tau )$ is the same as Eq.~(\ref{Uasy})
which implies that the symmetry potential is identical for the eMDYI and MID
interactions. Therefore, in the eMDYI interaction, the isoscalar part of the
single nucleon potential is momentum dependent but the nuclear symmetry
potential is not. For symmetric nuclear matter, the single nucleon potential
in the eMDYI interaction is exactly the same as that in the MDI interaction.
We note that the same strategy has been used to study the momentum
dependence effects in heavy-ion collisions in a previous work \cite{Che04}.

\subsection{Thermodynamic quantities of asymmetric nuclear matter}

At zero temperature, $f_{\tau }(\vec{r},\vec{p})$ $=\frac{2}{h^{3}}\Theta
(p_{f}(\tau )-p)$ and all the integrals in previous expressions can be
calculated analytically \cite{Che07}, while at a finite temperature $T$, the
phase space distribution function becomes the Fermi distribution
\begin{equation}
f_{\tau }(\vec{r},\vec{p})=\frac{2}{h^{3}}\frac{1}{\exp (\frac{\frac{p^{2}}{%
2m_{_{\tau }}}+U_{\tau }-\mu _{\tau }}{T})+1}  \label{f}
\end{equation}%
where $\mu _{\tau }$ is the proton or neutron chemical potential and can be
determined from%
\begin{equation}
\rho _{\tau }=\int f_{\tau }(\vec{r},\vec{p})d^{3}p.
\end{equation}%
In the above, $m_{_{\tau }}$ is the proton or neutron mass and $U_{\tau }$
is the proton or neutron single nucleon potential in different interactions.
For fixed density $\rho $, temperature $T$, and isospin asymmetry $\delta $,
the chemical potential $\mu _{\tau }$ and the distribution function $f_{\tau
}(\vec{r},\vec{p})$ can be determined numerically by a self-consistency
iteration scheme \cite{Gal90,Xu07}. From the chemical potential $\mu _{\tau
} $ and the distribution function $f_{\tau }(\vec{r},\vec{p})$, the energy
per nucleon $E(\rho ,T,\delta )$ can be obtained as
\begin{equation}
E(\rho ,T,\delta )=\frac{1}{\rho }\left[ V(\rho ,T,\delta )+{\sum_{\tau }}%
\int d^{3}p\frac{p^{2}}{2m_{\tau }}f_{\tau }(\vec{r},\vec{p})\right] .
\label{E}
\end{equation}%
Furthermore, we can obtain the entropy per nucleon $S_{\tau }(\rho ,T,\delta
)$ as
\begin{equation}
S_{\tau }(\rho ,T,\delta )=-\frac{8\pi }{{\rho }h^{3}}\int_{0}^{\infty
}p^{2}[n_{\tau }\ln n_{\tau }+(1-n_{\tau })\ln (1-n_{\tau })]dp  \label{S}
\end{equation}%
with the occupation probability%
\begin{equation}
n_{\tau }=\frac{1}{\exp (\frac{\frac{p^{2}}{2m_{_{\tau }}}+U_{\tau }-\mu
_{\tau }}{T})+1}.
\end{equation}%
Finally, the pressure $P(\rho ,T,\delta )$ can be calculated from the
thermodynamic relation
\begin{eqnarray}
P(\rho ,T,\delta ) &=&\left[ T{\sum_{\tau }}S_{\tau }(\rho ,T,\delta
)-E(\rho ,T,\delta )\right] \rho  \notag \\
&&+\sum_{\tau }\mu _{\tau }\rho _{\tau }.  \label{P}
\end{eqnarray}

\section{Thermal effects on the isospin-dependent bulk and single-particle
properties of asymmetric nuclear matter}

\label{bulk}

\subsection{Nuclear symmetry energy}

As in the case of zero temperature, phenomenological and microscopic studies
\cite{Che01b,Zuo03,Xu07,Mou07} indicate that the EOS of hot asymmetric
nuclear matter at density $\rho $, temperature $T$, and an isospin asymmetry
$\delta $ can also be written as a parabolic function of $\delta $, i.e.,
\begin{equation}
E(\rho ,T,\delta )=E(\rho ,T,\delta =0)+E_{sym}(\rho ,T)\delta ^{2}+\mathcal{%
O}(\delta ^{4}).  \label{eos}
\end{equation}%
This nice feature of the empirical parabolic law for the EOS of hot
asymmetric nuclear matter is very useful and convenient for the calculation
of the nuclear symmetry energy. With the empirical parabolic law, the
temperature and density dependent symmetry energy $E_{sym}(\rho ,T)$ for hot
asymmetric nuclear matter can thus be extracted from
\begin{equation}
E_{sym}(\rho ,T)\approx E(\rho ,T,\delta =1)-E(\rho ,T,\delta =0).
\end{equation}%
In this sense, the symmetry energy $E_{sym}(\rho ,T)$ gives estimation of
the energy cost to convert all protons in symmetry matter to neutrons at
fixed temperature $T$ and density $\rho $. The parabolic approximation has
been justified for MDI interaction in Ref.~\cite{Xu07}. We note that the
parabolic approximation also holds very well for the MID and eMDYI
interactions.
\begin{figure}[tbh]
\includegraphics[scale=0.8]{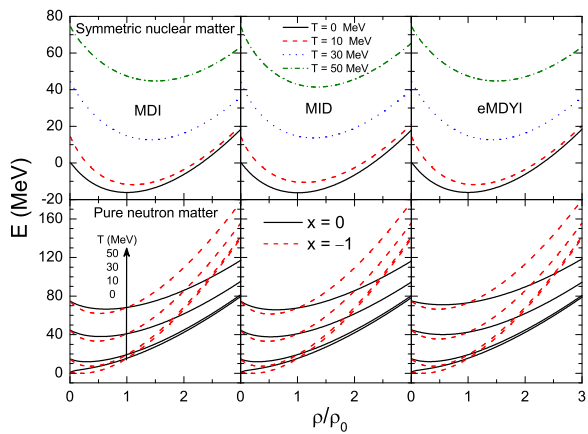}
\caption{{\protect\small (Color online) Density dependence of energy per
nucleon for symmetric nuclear matter (upper panels) and pure nuclear matter
(lower panels) in MDI, MID and eMDYI interactions at $T=0$, $10$, $30$ and $%
50$ MeV.}}
\label{EOS}
\end{figure}

Shown in Fig. \ref{EOS} is the density dependence of $E(\rho ,T,\delta )$
for symmetric nuclear matter and pure neutron matter at $T=0$, $10$, $30$
and $50$ MeV using the MDI, MID and eMDYI interactions with $x=0$ and $-1$.
For symmetric nuclear matter ($\delta =0$), the parameter $x=0$ would give
the same results as the parameter $x=$ $-1$ as we have discussed above, and
thus the curves shown in upper panels of Fig. \ref{EOS} are the same for $x=0
$ and $-1$. However, for pure neutron matter, the parameters $x=0$ and $-1$
display different density dependence for the energy per nucleon $E(\rho
,T,\delta )$, which just reflects that the parameters $x=0$ and $-1$ give
different density dependence of the nuclear symmetry energy as will be
discussed in the following. From Fig. \ref{EOS}, one can see that the energy
per nucleon $E(\rho ,T,\delta )$ increases with increasing temperature $T$.
The increment of the energy per nucleon $E(\rho ,T,\delta )$ with the
temperature reflects the thermal excitation of the nuclear matter due to the
change of the phase-space distribution function $f_{\tau }(\vec{r},\vec{p})$
or the occupation probability $n_{\tau }(\vec{r},\vec{p})$. With the
increment of the temperature, more nucleons move to higher momentum states
and thus lead to larger internal energy per nucleon. Furthermore, the
temperature effects are seen to be stronger at lower densities while they
become much weaker at higher densities. At lower densities, the Fermi
momentum $p_{f}(\tau )$ is smaller and thus temperature effects on the
energy per nucleon $E(\rho ,T,\delta )$ are expected to be stronger.
\begin{figure}[tbh]
\includegraphics[scale=1.2]{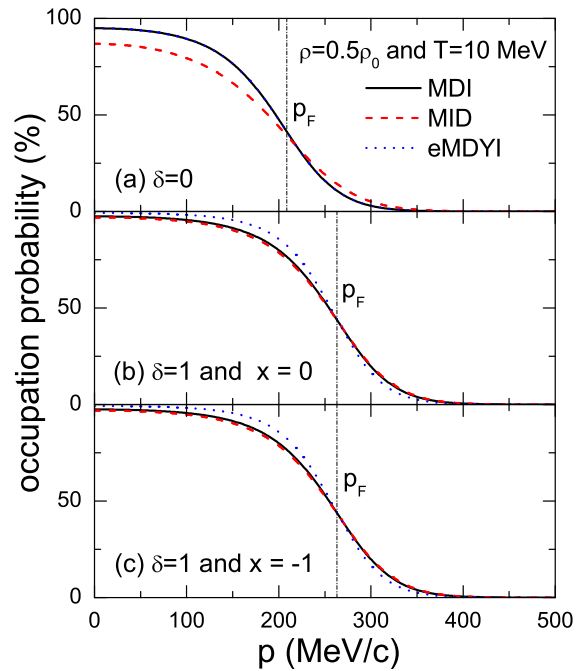}
\caption{{\protect\small (Color online) Occupation probability distribution
for symmetry nuclear matter and pure neutron matter at $\protect\rho =0.5%
\protect\rho _{0}$ and $T=10$ MeV in MDI, MID and eMDYI interactions with $%
x=0$ and $x=-1$}. The corresponding Fermi momentum $p_{f}(\protect\tau )$ at
zero temperature is indicated.}
\label{occ}
\end{figure}

In order to understand better the above results, we show in Fig.~\ref{occ}
the momentum dependence of the occupation probability $n_{\tau }(\vec{r},%
\vec{p})$ for symmetric nuclear matter and pure neutron matter at $\rho
=0.5\rho _{0}$ and $T=10$ MeV using the MDI, MID and eMDYI interactions with
$x=0$ and $-1$. For comparison, the corresponding Fermi momentum $p_{f}(\tau
)$ at zero temperature is also indicated. The occupation probability $%
n_{\tau }(\vec{r},\vec{p})$ at finite temperatures is self-consistently
determined and from which other quantities can be calculated. For symmetric
nuclear matter, as mentioned in Section \ref{model}, the MDI interaction has
the same occupation probability as the eMDYI interaction, and also the
occupation probability is independent of the $x$ parameter. One can see
clearly from Fig.~\ref{occ} that compared with the case of zero temperature,
more nucleons move to higher momentum states at finite temperature of $T=10$
MeV. In addition, the results indicate that for symmetric nuclear matter,
the occupation probability distribution of MID interaction is more extended
than those of MDI and eMDYI interactions. For pure neutron matter the result
of MDI interaction is very close to that of the MID interaction, while the
eMDYI interaction seems to have a steeper occupation probability
distribution. Meanwhile, the value of $x$ parameter seems to have little
effects on the shape of the occupation probability distribution even though
for pure neutron matter. Although here we only show the case of $\rho
=0.5\rho _{0}$, we note that these properties hold for all the densities.

In addition, it is seen from Fig. \ref{EOS} that the MDI, MID and eMDYI
interactions give almost the same EOS for cold nuclear matter even at high
densities though their parameters are constrained only at saturation density
as discussed in Section \ref{model}. Meanwhile, it is interesting to see
that the MDI, MID and eMDYI interactions also produce quite similar EOS at
finite temperatures, even at high densities, for both $x=0$ and $x=-1$. This
is due to the fact that only low momentum parts of the single-particle
potential is lifted at finite temperatures as we will show in the following
while these momentum parts do not contribute much to the total energy of
system through integration with respect to the momentum. Although the three
interactions have different occupation probability distributions at finite
temperatures, the potential energy makes self-consistently the balance and
leads to that they have the similar EOS for symmetric nuclear matter and
pure neutron matter at finite temperatures. This feature implies that the
interaction of the simple momentum-independent MID type is enough to
describe the EOS of asymmetric nuclear matter at finite temperatures.
However, as will be discussed in following, we note that the chemical
potential, which is a quite important quantity in the study of the
mechanical and chemical instabilities as well as liquid-gas phase transition
in hot asymmetric nuclear matter, can be significantly different for the
three kinds of interactions.
\begin{figure}[tbh]
\includegraphics[scale=0.8]{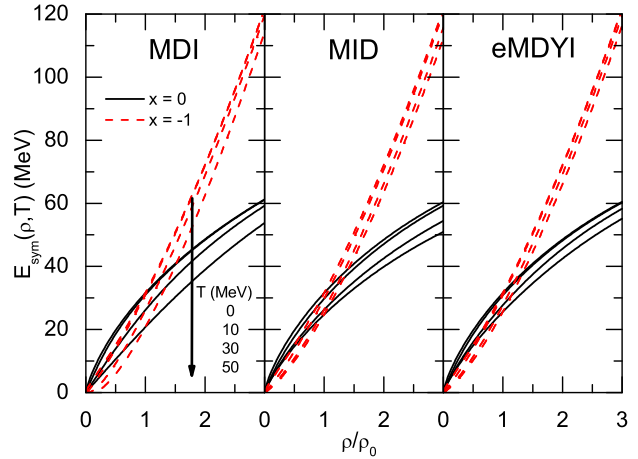}
\caption{{\protect\small (Color online) Density dependence of nuclear
symmetry energy in MDI, MID and eMDYI interactions with $x=0$ and $x=-1$ at $%
T=0$, $10$, $30$ and $50$ MeV.}}
\label{Esym}
\end{figure}

Now let's see the temperature dependence of the nuclear symmetry energy. In
Fig.~\ref{Esym} we show the density dependence of the nuclear symmetry
energy at $T=0$, $10$, $30$ and $50$ MeV using the MDI, MID and eMDYI
interactions with $x=0$ and $-1$. For different choice of the parameter $x=0$
and $-1$, $E_{sym}(\rho ,T)$ display different density dependence with $x=0$
($-1$) giving larger (smaller) values for the symmetry energy at lower
densities while smaller (larger) ones at higher densities for a fixed
temperature. For all the three interactions with both $x=0$ and $-1$, it is
seen that the symmetry energy decreases with increasing temperature. At
higher temperatures, one expects the symmetry energy $E_{sym}(\rho ,T)$ to
decrease as the Pauli blocking (a pure quantum effect) becomes less
important when the nucleon Fermi surfaces become more diffused at
increasingly higher temperatures \cite{Che01b,Zuo03,LiBA06c,Xu07}.

\begin{figure}[tbh]
\includegraphics[scale=0.85]{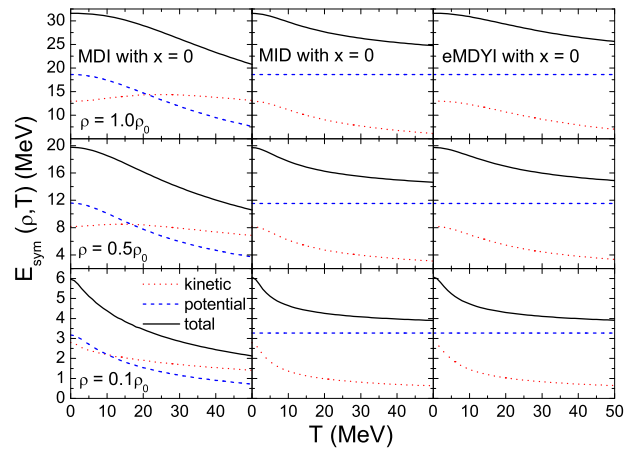}
\caption{{\protect\small (Color online) Temperature dependence of the total
symmetry energy and its kinetic and potential contributions in MDI, MID and
eMDYI interactions with $x=0$ at $\protect\rho =0.1\protect\rho _{0}$, $0.5%
\protect\rho _{0}$ and $1.0\protect\rho_{0} $.}}
\label{EsymT_x0}
\end{figure}

Within the present self-consistent thermal model, because the single
particle potential is isospin and momentum dependent with the MDI
interaction, the potential part of the symmetry energy with the MDI
interaction is thus temperature dependent as shown in Eq. (\ref{MDIV}). On
the other hand, the potential part of the symmetry energy with the MID and
eMDYI interactions does not depend on the temperature by the construction as
seen in Eq. (\ref{MIDV}) and Eq. (\ref{MDYIV}). It is thus interesting to
study how the potential and kinetic parts of the symmetry energy $%
E_{sym}(\rho ,T)$ may vary respectively with temperature, which will reflect
the effects of isospin and momentum dependence of the nuclear interaction.
Fig.~\ref{EsymT_x0} displays the temperature dependence of the symmetry
energy $E_{sym}(\rho ,T)$ as well as its potential and kinetic energy parts
using the MDI, MID and eMDYI interactions with $x=0$ at $\rho =1.0\rho _{0}$%
, $0.5\rho _{0}$, and $0.1\rho _{0}$. With the parameter $x=-1$, we note the
same conclusion is obtained. For the MDI interaction, it is seen that both
the total symmetry energy $E_{sym}(\rho ,T)$ and its potential energy part
decrease with increasing temperature at all three densities considered.
Meanwhile, one can see that the kinetic contribution increases slightly with
increasing temperature at low temperature and then decreases with increasing
temperature at high temperature for $\rho =1.0\rho _{0}$ and $0.5\rho _{0}$,
while it decreases monotonically for $\rho =0.1\rho _{0}$. These features
observed for the MDI interaction are uniquely determined by the isospin and
momentum dependence in the MDI interaction within the present
self-consistent thermal model. On the other hand, for MID and eMDYI
interactions the kinetic part of the total symmetry energy decreases with
increasing temperature at all the densities while the potential contribution
is independent of temperature and it has the same value for the MID and
eMDYI interactions. These features indicate that the temperature dependence
of the total symmetry energy is due to both the potential contribution and
kinetic contribution for MDI interaction, but it is only due to the kinetic
contribution for the MID and eMDYI interactions.

It should be mentioned that for the MDI interaction, the decrement
of the kinetic energy part of the symmetry energy with temperature
at very low densities is consistent with predictions of the free
Fermi gas model at high temperatures and/or very low densities
\cite{LiBA06c,Lee01,Mek05,Mou07}. Interestingly, we can see that
the temperature dependence of the total symmetry energy
$E_{sym}(\rho ,T)$ is quite similar for all the three interactions
except that the MDI interaction exhibits a little stronger
temperature dependence at higher temperatures. This is due to the
fact that the phase space distribution function will vary
self-consistently according to whether the single particle
potential is momentum dependent or not. In addition, we note that
as shown in Ref. \cite{Mou07}, both the potential and kinetic
parts of the symmetry energy $E_{sym}(\rho ,T)$ can decrease with
temperature for all the densities considered there by using the
isospin- and momentum-dependent BGBD interaction developed by
Bombaci \cite{Bom01} based on the well known Gale-Bertsch-Das
Gupta formalism \cite{Gal87}. The different temperature dependence
of the potential and kinetic parts of the symmetry energy between
the MDI and BGBD interaction is due to the fact that the MDI and
BGBD interactions have different forms of the energy density
functional and the MDI interaction leads to a more complicated
momentum dependence of the single-particle potential. This feature
implies that the temperature dependence of the potential and
kinetic parts of the symmetry energy depends on the isospin and
momentum dependence of the nuclear interactions.

\subsection{Nuclear symmetry potential}

To understand more clearly the effects of the isospin and momentum
dependence of nuclear interactions, we discuss the single particle potential
and its temperature dependence with the three models. In particular, we
study the thermal effects on the nuclear symmetry potential. The nuclear
symmetry potential refers to the isovector part of the nucleon mean-field
potential in isospin asymmetric nuclear matter. Besides the nuclear density,
the symmetry potential of a nucleon in nuclear matter also depends on the
momentum or energy of the nucleon. In hot asymmetric nuclear matter, the
symmetry potential of a nucleon can also depend on the temperature. The
nuclear symmetry potential is different from the nuclear symmetry energy as
the latter involves the integration of the isospin-dependent mean-field
potential of a nucleon over its momentum. Both the nuclear symmetry
potential and the nuclear symmetry energy are important for understanding
many physics questions in nuclear physics and astrophysics. Various
microscopic and phenomenological models have been used to study the symmetry
potential \cite%
{Ulr97,Fuc04,Ma04,Sam05a,Fuc05,Fuc05b,Ron06,Bom91,Zuo05,Bar05,Das03,LiBA04a,LiBA04c,Che04,Riz04,Beh05,Riz05}%
, and the predicted results vary widely as in the case of the nuclear
symmetry energy. In particular, whereas most models predict a decreasing
symmetry potential with increasing nucleon momentum albeit at different
rates, a few nuclear effective interactions used in some models give an
opposite behavior. However, all the above studies on the nuclear symmetry
potential are for zero-temperature and the temperature dependence of the
nuclear symmetry potential has received so far little theoretical attention.
The density, temperature and momentum dependent nuclear symmetry potential
can be evaluated from
\begin{equation}
U_{\mathrm{sym}}(\rho ,\vec{p},T)=\frac{U_{n}(\rho ,\vec{p},T)-U_{p}(\rho ,%
\vec{p},T)}{2\delta }  \label{Usym}
\end{equation}%
where $U_{n}(\rho ,\vec{p},T)$ and $U_{p}(\rho ,\vec{p},T)$
represent, respectively, the neutron and proton single-particle
potentials in hot asymmetric nuclear matter. In the present work,
we use $\delta =0.5$ in the calculation for the symmetry potential
and we note that the result can vary within only several percents
at most by choosing different values of $\delta $.

\begin{figure}[tbh]
\includegraphics[scale=0.8]{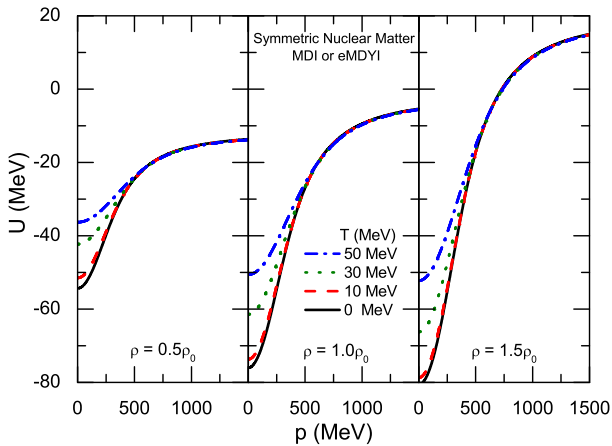}
\caption{{\protect\small (Color online) Momentum dependence of the single
particle potential in symmetric nuclear matter at $\protect\rho= 0.5\protect%
\rho _{0}$, $1.0\protect\rho_{0}$ and $1.5\protect\rho _{0}$ and $T=0$, $10$%
, $30$ and $50$ MeV in the MDI or eMDYI interaction.}}
\label{SPPd0}
\end{figure}

In order to see the temperature effects on the nuclear symmetry
potential, we first study the temperature dependence of the
nucleon single-particle potential in hot nuclear matter. In
Fig.~\ref{SPPd0} we show the momentum dependence of the single
particle potential of symmetric nuclear matter at $T=0$, $10$,
$30$ and $50$ MeV and $\rho =$ $0.5\rho _{0}$, $1.0\rho _{0}$ and
$1.5\rho _{0}$ for the MDI or eMDYI interaction. As mentioned in
Section \ref{model}, for symmetric nuclear matter the MDI and
eMDYI interactions are exactly the same while the MID interaction
is completely momentum-independent and thus
temperature-independent as well. Therefore, we do not discuss here
the single particle potential for the MID interaction. From
Fig.~\ref{SPPd0}, it is seen that the single particle potentials
increase with increasing momentum and saturate at high momenta. In
addition, the shape of single-particle potentials are steeper with
increasing density and so the momentum dependence becomes stronger
at higher densities. It is interesting to see that only low
momentum parts of the potentials are lifted with increasing
temperature, and this feature is quite reasonable and expected
since the nucleons with high momenta can hardly be affected by the
temperature.

\begin{figure}[tbh]
\includegraphics[scale=0.8]{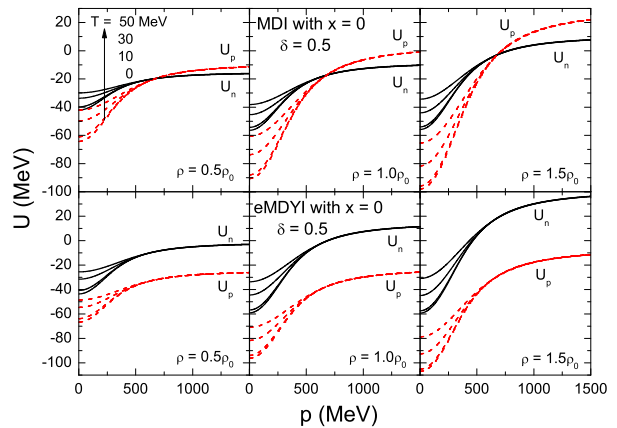}
\caption{{\protect\small (Color online) Momentum dependence of the single
particle potential of protons and neutrons in asymmetric nuclear matter with
isospin asymmetry $\protect\delta =0.5$ at $\protect\rho= 0.5\protect\rho %
_{0}$, $1.0\protect\rho _{0}$ and $1.5\protect\rho _{0}$ and $T=0$, $10$, $30
$ and $50$ MeV in the MDI and eMDYI interactions with $x=0$.}}
\label{SPPd05x0}
\end{figure}

\begin{figure}[tbh]
\includegraphics[scale=0.8]{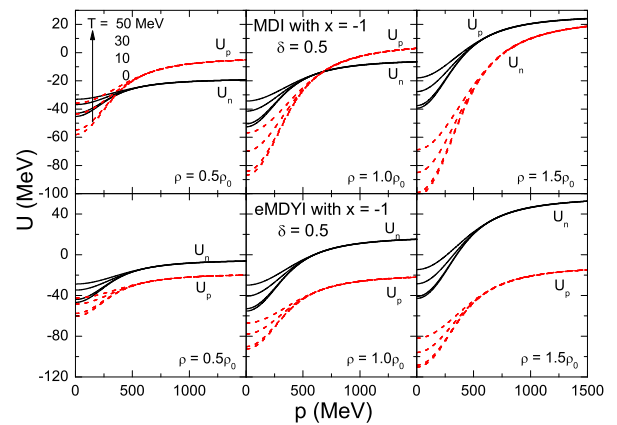}
\caption{{\protect\small (Color online) Same as Fig.~\protect\ref{SPPd05x0}
but for $x=-1$.}}
\label{SPPd05xm1}
\end{figure}

Shown in Fig.~\ref{SPPd05x0} is the momentum dependence of single particle
potentials of protons and neutrons in asymmetric nuclear matter with the
isospin asymmetry of $\delta =0.5$ at $T=0$, $10$, $30$ and $50$ MeV and $%
\rho =0.5\rho _{0}$, $1.0\rho _{0}$ and $1.5\rho _{0}$ for MDI (upper
panels) and eMDYI (lower panels) interactions with $x=0$. The similar
results are shown in Fig.~\ref{SPPd05xm1} for $x=-1$. The temperature and
density effects are seen to be very similar to that in the case of symmetric
nuclear matter as shown in Fig.~\ref{SPPd0}, i.e., only low momentum parts
of the potentials are lifted with increasing temperature. In contrast to the
results in the case of symmetric nuclear matter, however, the neutron and
proton single particle potentials in asymmetric nuclear matter at a fixed
temperature becomes different from each other. For the eMDYI interaction,
the potentials at a fixed temperature are just momentum-independently
shifted for protons and neutrons with a higher potential for neutrons while
lower potential for protons, and the shifted value is sensitive to the
density and the $x$ value. For the MDI interaction, the isospin and momentum
dependence of single particle potentials is somewhat complicated. At a fixed
temperature, for MDI interaction with $x=0$, the potential of neutrons is
larger than that of protons at low momenta but the opposite result is
obtained at high momenta, which indicates a steeper potential (stronger
momentum dependence) for protons than for neutrons. For different $x$
values, the single particle potentials of MDI interaction are also seen to
be shifted and the shifted value depending on the density and this is due to
the fact that the term with $x$ in Eq.~( \ref{MDIU}) is momentum-independent
and depends only on the density.

\begin{figure}[tbh]
\includegraphics[scale=0.8]{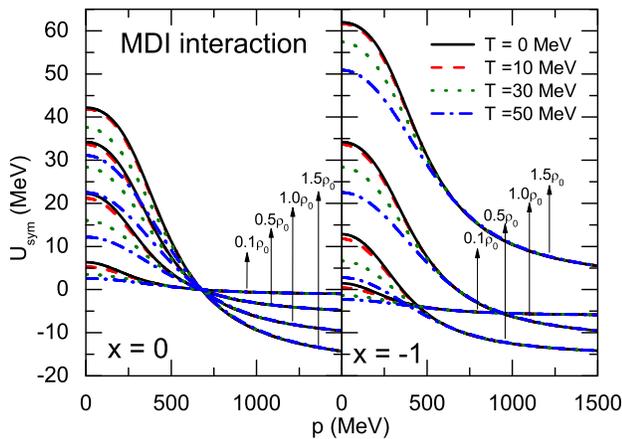}
\caption{{\protect\small (Color online) Momentum dependence of the symmetry
potential at $\protect\rho= 0.1\protect\rho _{0}$, $0.5\protect\rho _{0}$, $%
1.0\protect\rho _{0}$ and $1.5\protect\rho _{0}$ and $T=0$, $10$, $30$ and $%
50$ MeV in the MDI interaction with $x=0$ and $x=-1$.}}
\label{Usymmdi}
\end{figure}

The above discussions imply that the nuclear symmetry potential for the
eMDYI interaction does not depend on the momentum while that for the MDI
interaction does. Show in Fig.~\ref{Usymmdi} is the momentum dependence of
the nuclear symmetry potential at $T=0$, $10$, $30$ and $50$ MeV and $\rho
=0.1\rho _{0}$, $0.5\rho _{0}$, $1.0\rho _{0}$ and $1.5\rho _{0}$ using the
MDI interaction with $x=0$ and $x=-1$. It is seen that the symmetry
potential decreases with increasing momentum for both $x=0$ and $x=-1$.
Empirically, for the momentum dependence of the nuclear symmetry potential
at zero temperature, a systematic analysis of a large number of
nucleon-nucleus scattering experiments and (p,n) charge-exchange reactions
at beam energies up to about $100$ MeV has shown that the data can be very
well described by the parametrization $U_{\mathrm{sym}}=a-bE_{\mathrm{kin}}$
with $a\approx 22-34$ MeV and $b\approx 0.1-0.2$ \cite%
{Sat69,Hof72,Hod94,Kon03}. Although the uncertainties in both
parameters $a$ and $b$ are large, the nuclear symmetry potential
at saturation density, i.e., the Lane potential
$U_{\mathrm{Lane}}$ \cite{Lan62}, clearly decreases approximately
linearly with increasing beam energy. This provides a stringent
constraint on the low energy behavior of the nuclear symmetry
potential at saturation density. We note that for the MDI
interaction with both $x=0$ and $x=-1$ at saturation density, the
symmetry potential agrees very well with the empirical Lane
potential. In addition, the high energy behavior of the nuclear
symmetry potential from the MDI interaction with $x=0 $ has been
also found to be consistent with the results from recent study
\cite{Che05c,LiZH06b} in the relativistic impulse ($t$-$\rho $)
approximation based on the empirical NN scattering amplitude
\cite{Mcn83b} or the Love-Franey NN scattering amplitude developed
by Murdock and Horowitz \cite{Hor85,Mur87}. On the other hand, the
low energy behavior of the nuclear symmetry potential at densities
away from normal nuclear density is presently not known
empirically. Experimental determination of both the density and
momentum dependence of the nuclear symmetry potential is thus of
great interest, and heavy-ion reactions with radioactive beams
provide a unique tool to extract this information in terrestrial
laboratories \cite{LiBA04a,Riz05,LiBA07a}. From
Fig.~\ref{Usymmdi}, one can see clearly that the symmetry
potentials decrease with increasing temperature, especially at low
momenta. At high momenta (above about $500$ MeV/c), the
temperature effect on the symmetry potential is quite weak due to
the fact that the nucleons with high momenta can hardly be
affected by the temperature as mentioned above.

\subsection{Isospin-splitting of nucleon effective mass}

One of the important single-particle properties of nuclear matter is the
nucleon effective mass which characterizes the momentum dependence of the
single-particle potential of a nucleon. The nucleon effective mass $m_{\tau
}^{\ast }$ is defined by%
\begin{equation}
\frac{m_{\tau }^{\ast }}{m_{\tau }}=\frac{p}{m_{\tau }\frac{d\epsilon _{\tau
}}{dp}}=\left( 1+\frac{m_{\tau }}{p}\frac{dU_{\tau }}{dp}\right) ^{-1}.
\label{Meff}
\end{equation}%
where $\epsilon _{\tau }$ represents the single-particle energy while $%
U_{\tau }$ is the single-particle potential. In such a way, the nucleon
effective mass $m_{\tau }^{\ast }$ is related to the density of states $%
m_{\tau }^{\ast }/(2\pi \hbar )^{3}$ in asymmetric nuclear matter. By
definition, the nucleon effective mass generally depends on the density,
isospin asymmetry of the medium, and the momentum of the nucleon \cite%
{Fuc05,Jam89,Neg98}. In hot nuclear medium, it depends on the temperature as
well. At zero temperature, when the nucleon effective mass is evaluated at
the Fermi momentum $p_{\tau }=p_{f}({\tau })$, Eq. (\ref{Meff}) yields the
Landau mass which is related to the $f_{1}$ Landau parameter of a Fermi
liquid \cite{Fuc05,Jam89,Neg98}. A detailed discussion about different kinds
of effective masses can be found in Ref. \cite{Jam89}.

With the single-particle potential in Eq. (\ref{MDIU}), since the
momentum-dependent part of the nuclear potential is independent of the
parameter $x$, the nucleon effective masses are independent of the $x$
parameter too. The neutron and proton effective masses are usually different
in asymmetric nuclear matter due to the fact that the momentum dependence of
the single-particle potential is different for neutrons and protons in
asymmetric nuclear matter. The isospin-splitting of nucleon effective mass
in asymmetric nuclear matter, i.e., the difference between the neutron and
proton effective masses is currently not known empirically \cite{Lun03}. We
note that theoretically the neutron-proton effective mass splitting is still
highly controversial within different approaches and/or using different
nuclear effective interactions \cite{Riz04,LiBA04c,Beh05,Che07}. Being
phenomenological and non-relativistic in nature, the neutron-proton
effective mass splitting in the MDI interaction is consistent with
predictions of all non-relativistic microscopic models, see, e.g., \cite%
{Bom91,Zuo05,Sjo76}, and the non-relativistic limit of microscopic
relativistic many-body theories, see, e.g., Refs. \cite%
{Sam05a,Ma04,Fuc04,Fuc05}. Recent transport model studies indicate that the
neutron/proton ratio at high transverse momenta and/or rapidities is a
potentially useful probe of the neutron-proton effective mass splitting in
neutron-rich matter \cite{LiBA04a,Riz05}. Since the momentum dependence of
the single-particle potential is usually temperature dependent, it is thus
interesting to see the temperature effect on the nucleon effective mass.
\begin{figure}[tbh]
\includegraphics[scale=0.8]{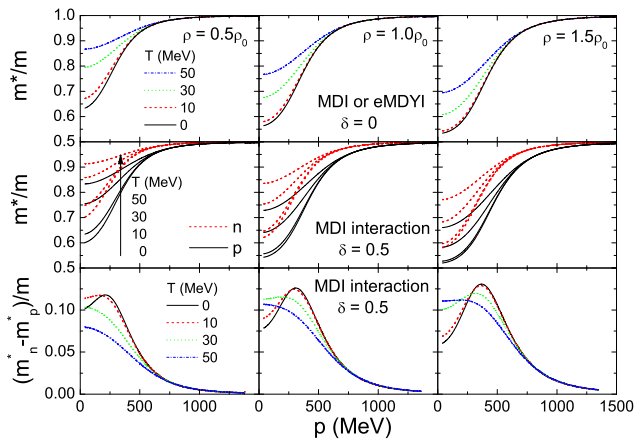}
\caption{{\protect\small (Color online) Momentum dependence of the effective
mass of protons and neutrons in symmetric nuclear matter ($\protect\delta =0$%
, upper panels) and asymmetric nuclear matter ($\protect\delta =0.5$, middle
panels) at $\protect\rho= 0.5\protect\rho _{0}$, $1.0\protect\rho _{0}$ and $%
1.5\protect\rho _{0}$ and $T=0$, $10$, $30$ and $50$ MeV in the MDI and
eMDYI interactions. The corresponding results for the reduced
isospin-splitting of the nucleon effective mass, i.e., $(m_{n}^{\ast
}-m_{p}^{\ast })/m$ in asymmetric nuclear matter ($\protect\delta =0.5$) are
also shown in the lower panels.}}
\label{EffMass}
\end{figure}

In the upper panels of Fig.~\ref{EffMass}, we show the momentum dependence
of nucleon effective mass in symmetric nuclear matter at $\rho =0.5\rho _{0}$%
, $1.0\rho _{0}$ and $1.5\rho _{0}$ and $T=0$, $10$, $30$ and $50$ MeV for
the MDI or eMDYI interaction. The similar results for the neutron and proton
effective masses in neutron-rich nuclear matter with isospin asymmetry $%
\delta =0.5$ are displayed in the middle panels of Fig.~\ref{EffMass} with
the MDI interaction. It should be noted that for the symmetric nuclear
matter, the eMDYI interaction gives the same nucleon effective mass as the
MDI interaction. In addition, the effective mass is independent of the value
of $x$ since the the momentum dependence of the single particle potential
does not depend on the $x$ parameter as mentioned above. For the MID
interaction, the single particle potential is completely
momentum-independent and thus temperature-independent obviously, and the
nucleon effective mass is just equal to the nucleon mass in free space. For
a fixed temperature, one can see clearly from the upper and middle panels of
Fig.~\ref{EffMass} that the nucleon effective mass decreases with increasing
density and decreasing momentum, which indicates that the momentum
dependence of single-particle potential is stronger at higher densities and
lower momenta. At high momentum, the nucleon effective mass approaches to
the nucleon mass in free space as the single particle potential is
saturated. For a fixed momentum, the nucleon effective mass increases with
temperature, especially at lower momenta, which indicates that the
temperature effect weakens the momentum dependence of the nuclear
interaction at lower momenta. In asymmetric nuclear matter at a fixed
temperature, the neutron effective mass is seen to be larger than proton
effective mass at a fixed momentum, and thus leads to the isospin-splitting
of the nucleon effective mass.

In order to see the temperature effect on the isospin-splitting of the
nucleon effective mass, we show the corresponding momentum dependence of the
reduced isospin-splitting of the nucleon effective mass, i.e., $(m_{n}^{\ast
}-m_{p}^{\ast })/m$ in the lower panels of Fig.~\ref{EffMass}. It is
indicated that the temperature effect on the isospin-splitting of the
nucleon effective mass displays some complicated behaviors. At lower
densities, the temperature effect seems to reduce the isospin-splitting of
the nucleon effective mass for a fixed momentum. However, at higher
densities, it depends on the momentum, i.e., at higher momenta, the
temperature effect reduces the isospin-splitting of the nucleon effective
mass while it increases the isospin-splitting at lower momenta. These
features reflect the complexity of the temperature effects on the momentum
dependence of the neutron and proton single-particle potential in hot
asymmetric nuclear matter for the MDI interaction. It should be mentioned
that, at zero temperature, the nucleon effective mass is usually calculated
at the Fermi momentum $p_{\tau }=p_{f}({\tau })$, and thus the
isospin-splitting is generally a function of the density and the isospin
asymmetry.

\section{Mechanical and chemical instabilities}

\label{instability}

The mechanical and chemical instabilities of hot asymmetric nuclear matter
have been extensively studied based on various theoretical models \cite%
{Lat78,Bar80,Mul95,LiBA97b,Bar98,Bar01,Cat01,LiBA02b,Lee07}. However,
effects of the momentum-dependent interactions on the mechanical and
chemical instabilities have received so far little theoretical attention. In
the following, we study the mechanical and chemical instabilities using the
MDI, MID, and eMDYI interactions and mainly focus on the effects of the
isospin and momentum dependence of nuclear interactions.

\subsection{Mechanical instability}

The mechanical stability condition for a hot asymmetric nuclear matter is
\begin{equation}
\left( \frac{\partial P}{\partial \rho }\right) _{T,\delta }\geq 0.
\label{Mstability}
\end{equation}%
If the above condition is not satisfied, any growth in density
leads to the decrement of pressure. As the pressure is lower than
its background, the nuclear matter will be compacted, which leads
to the further growth of the density. In such way, any small
density fluctuations can grow and the nuclear matter becomes
mechanically unstable.

\begin{figure}[tbh]
\includegraphics[scale=0.8]{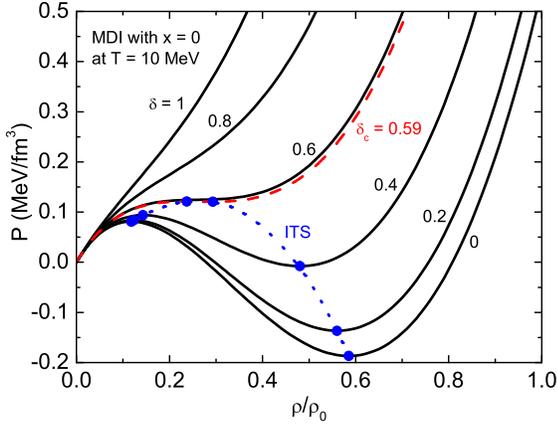}
\caption{{\protect\small (Color online) Pressure as a function of density at
fixed isospin asymmetry in the MDI interaction with $x=0$ at $T=10$ MeV. The
isothermal spinodals (ITS) and the case of critical isospin asymmetry are
also indicated.}}
\label{example_ITS1}
\end{figure}

As an example to see the picture of the boundary of mechanical
instability and the critical case, we show in
Fig.~\ref{example_ITS1} the isothermal line in the $P\sim \rho $
plane for asymmetric nuclear matter at $T=10$ MeV and different
values of the isospin asymmetry $\delta $ using the MDI
interaction with $x=0$. For smaller isospin asymmetries, the
mechanical stability condition can be violated (below the dotted
line). When $\delta $ is higher than the critical value (about
$0.59$ in this case), the pressure increases monotonically with
density and Eq.~(\ref{Mstability}) is satisfied for all densities.
In the critical case, we have
\begin{equation}
\left( \frac{\partial P}{\partial \rho }\right) _{T,\delta _{c}}=\left(
\frac{\partial ^{2}P}{\partial \rho ^{2}}\right) _{T,\delta _{c}}=0,
\label{inflection1}
\end{equation}%
where $\delta _{c}$ is the critical isospin asymmetry. Below the critical
isospin asymmetry, the extrema of the $P\sim \rho $ curve at different
isospin asymmetries form the boundary of the mechanical instability region,
namely, the isothermal spinodal (ITS).
\begin{figure}[tbh]
\includegraphics[scale=0.8]{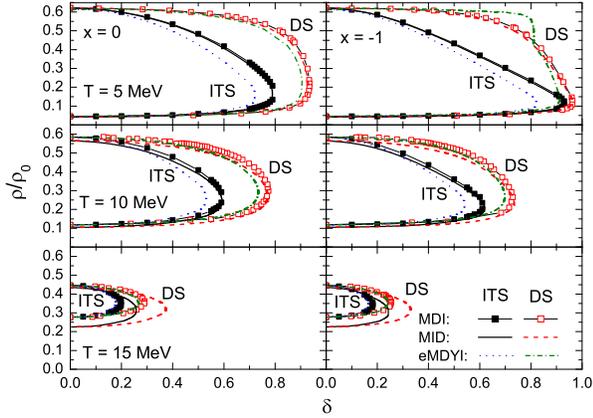}
\caption{{\protect\small (Color online) The boundary of mechanical (ITS) and
chemical (DS) instabilities in the $\protect\rho \sim \protect\delta $ plane
at $T=5$, $10$ and $15$ MeV for the MDI, MID and eMDYI interactions with $x=0
$ and $x=-1$.}}
\label{rhodelta1}
\end{figure}

\begin{figure}[tbh]
\includegraphics[scale=0.8]{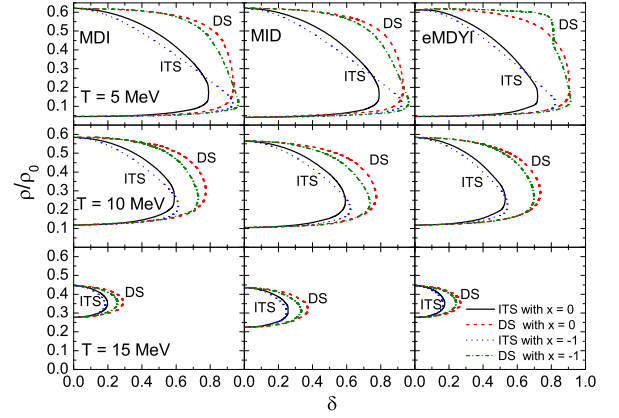}
\caption{{\protect\small (Color online) Same as Fig.~\protect\ref{rhodelta1}
but separately for the MDI, MID and eMDYI interactions to show the $x$
dependence.}}
\label{rhodelta2}
\end{figure}
\begin{figure}[tbh]
\includegraphics[scale=0.8]{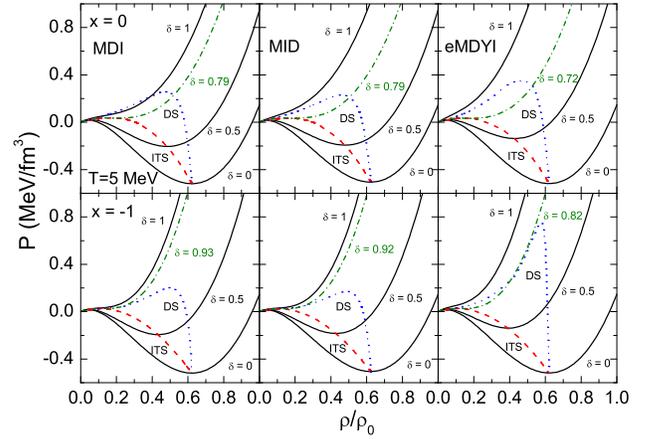}
\caption{{\protect\small (Color online) The boundary of mechanical (ITS) and
chemical (DS) instabilities in the $P\sim \protect\rho $ plane at $T=5$ MeV
for the MDI, MID and eMDYI interactions with $x=0$ and $x=-1$.}}
\label{Prho_T5}
\end{figure}

\begin{figure}[tbh]
\includegraphics[scale=0.8]{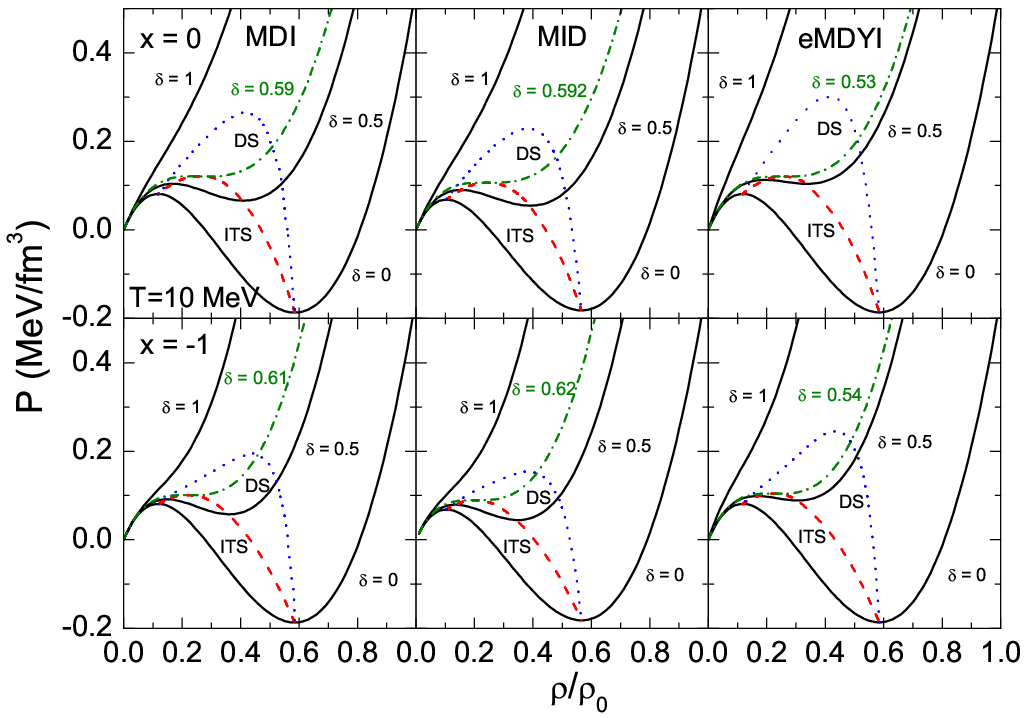}
\caption{{\protect\small (Color online) Same as Fig.~\protect\ref{Prho_T5}
but for $T=10$ MeV.}}
\label{Prho_T10}
\end{figure}

\begin{figure}[tbh]
\includegraphics[scale=0.8]{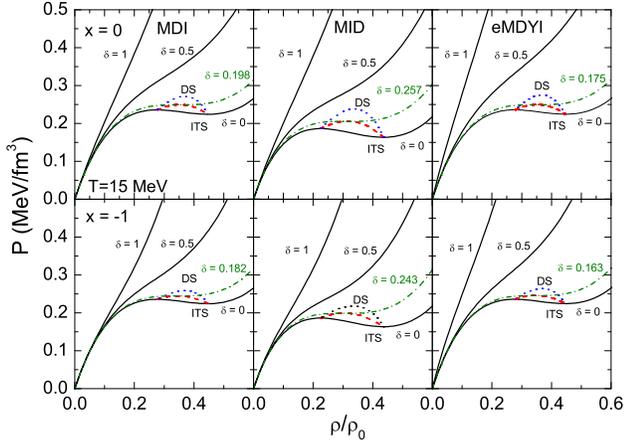}
\caption{{\protect\small (Color online) Same as Fig.~\protect\ref{Prho_T5}
but for $T=15$ MeV.}}
\label{Prho_T15}
\end{figure}

Figs.~\ref{rhodelta1} and \ref{rhodelta2} display the boundary of mechanical
instability, i.e., ITS for the MDI, MID and eMDYI interactions at $T=5,10$
and $15$ MeV with $x=0$ and $x=-1$ in the $\rho \sim \delta $ plane while
Figs.~\ref{Prho_T5}, \ref{Prho_T10} and \ref{Prho_T15} display the same
curves in the $P\sim \rho $ plane with the inclusion of the curves at
constant isospin asymmetries of $\delta =0$, $0.5$, $1$ and $\delta _{c}$.
From Fig.~\ref{rhodelta1} and \ref{rhodelta2}, one sees that nuclear matter
in the left region of the boundary of mechanical instability indicated by
ITS is mechanically unstable, and the critical isospin asymmetry as well as
the area of the mechanical instability region decreases with increasing
temperature. For each interaction, the boundaries overlap at $\delta =0$ for
different values of $x$, since for symmetry nuclear matter the three
interactions are independent of the value of $x$. For the MDI and eMDYI
interactions, the ITS has the same value at $\delta =0 $, as for symmetric
nuclear matter they are exactly the same model as mentioned in Section \ref%
{model}, while for MID interaction it is shifted to smaller densities at $%
\delta =0$. The critical isospin asymmetry is seen to be sensitive
to the density dependence of the symmetry energy, which can be
seen more clearly from Figs.~\ref{Prho_T5}, \ref{Prho_T10} and
\ref{Prho_T15} where the value of $\delta _{c}$ is indicated
exactly by the dash-dotted lines. At $T=5$ and $10$ MeV the
critical isospin asymmetry is larger for $x=-1$ than for $x=0$,
while at $T=15$ MeV it is smaller for $x=-1$ than $x=0$. These
phenomena indicate that the density dependence of nuclear symmetry
energy and the temperature are two important factors to determine
the critical isospin asymmetry and the area
of mechanical instability. In addition, it is seen from Figs.~\ref%
{rhodelta1} and \ref{rhodelta2} that the critical isospin
asymmetry and the area of mechanical instability are also
sensitive to the isospin and momentum dependence of the nuclear
interaction, especially at higher temperatures. Detailed
comparison indicates that the critical isospin asymmetry of the
MDI interaction is very similar to that of the MID interaction at
low and moderate temperatures, while it is similar to that of the
eMDYI interaction at high temperatures. Meanwhile, the area of the
mechanically unstable region is seen to be the largest for the MID
interaction while the smallest for the eMDYI interaction.
\begin{figure}[tbh]
\includegraphics[scale=0.8]{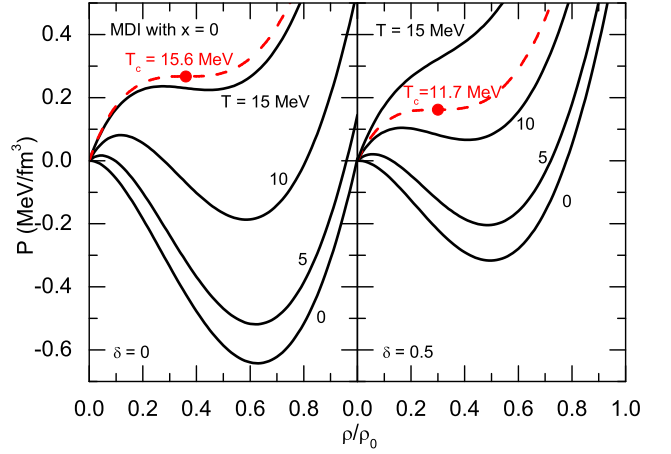}
\caption{{\protect\small (Color online) Pressure as a function of density at
fixed temperature in the MDI interaction with $x=0$ at $\protect\delta =0.0$
(left panel) and $\protect\delta =0.5$ (right panel). The case of critical
temperature is also indicated.}}
\label{example_ITS2}
\end{figure}

\begin{figure}[tbh]
\includegraphics[scale=0.8]{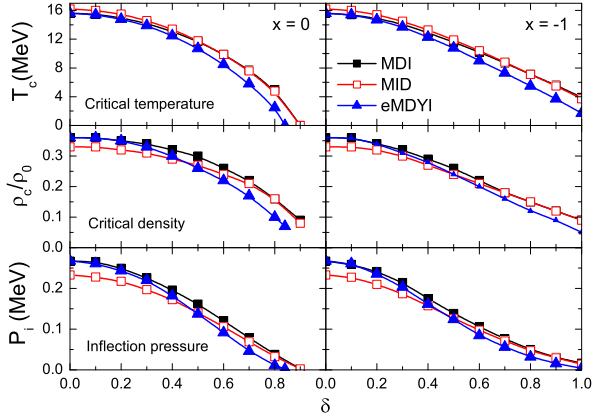}
\caption{{\protect\small (Color online) Isospin asymmetry
dependence of the critical temperature (upper panels), the
critical density (middle panels), and the inflection pressure
(lower panels) in the MDI, MID and eMDYI interactions with $x=0$
and $x=-1$.}} \label{deltaTcrhoc1}
\end{figure}

In the above calculations, we fixed the temperature while varied the isospin
asymmetry and thus got the critical isospin asymmetry. In Fig.~\ref%
{example_ITS2} we fix the isospin asymmetry at $\delta =0$ and $\delta =0.5$%
, respectively, and change the temperature. The case of MDI
interaction with $x=0$ is shown as an example. It is clearly shown
that increasing the temperature at fixed isospin asymmetry is just
like increasing the isospin asymmetry at fixed temperature, and
the mechanical stability condition Eq.~(\ref{Mstability}) is
satisfied in all the densities when the temperature is larger than
a critical temperature $T_{c}$ (about $15.6$ MeV at $\delta =0$
and $11.7$ MeV at $\delta =0.5$). The density at the inflection
point, which satisfies
\begin{equation}
\left( \frac{\partial P}{\partial \rho }\right) _{T_{c},\delta }=\left(
\frac{\partial ^{2}P}{\partial \rho ^{2}}\right) _{T_{c},\delta }=0,
\label{inflection2}
\end{equation}%
is the critical density $\rho _{c}$, and the pressure at the
inflection point is named after as the inflection pressure
$P_{i}$. In Fig.~\ref{deltaTcrhoc1}, we show the isospin asymmetry
dependence of the critical temperature $T_{c}$, the critical
density $\rho _{c}$, the inflection pressure $P_{i}$ for the MDI,
MID and eMDYI interactions with $x=0$ and $x=-1 $. It is indicated
that the critical temperature, critical density and the inflection
pressure decrease with increasing isospin asymmetry. Below the
curves the system can be mechanically unstable. The critical
temperature for symmetric nuclear matter is $15.6$ MeV for the MDI
and eMDYI interactions and $16.2$ MeV for the MID interaction with
both $x=0$ and $x=-1$. For $x=0$ the system is stable above a
certain high isospin asymmetry ($0.9$ for MDI and MID model and
$0.84$ for eMDYI model), but for $x=-1$ it can be mechanically
unstable even for pure neutron matter. These features indicate
again that the boundary of mechanical instability is quit
sensitive to the value of $x$. Furthermore, it is clearly seen
from Fig.~\ref{deltaTcrhoc1} that the MDI interaction is similar
to the MID interaction at low temperatures, while it is similar to
the eMDYI interaction at high temperatures.
\begin{figure}[tbh]
\includegraphics[scale=0.8]{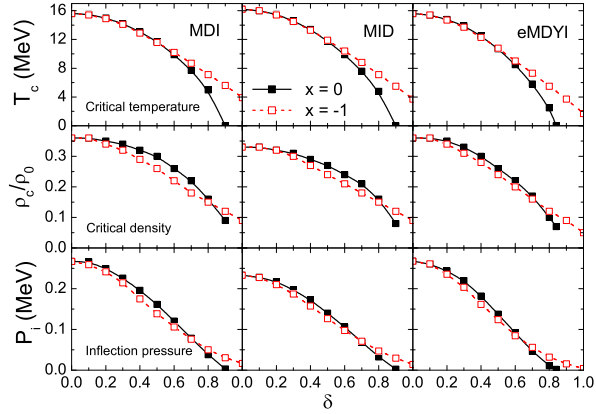}
\caption{{\protect\small (Color online) Same as Fig.~\protect\ref%
{deltaTcrhoc1} but separately for the MDI, MID and eMDYI interactions with $%
x=0$ and $x=-1$.}}
\label{deltaTcrhoc2}
\end{figure}

To see more clearly the effect of the density dependence of the symmetry
energy on the critical temperature, critical density and inflection
pressure, we show separately their isospin asymmetry dependence for each
interaction with $x=0$ and $x=-1$ in Fig.~\ref{deltaTcrhoc2}. In each case
the critical temperature for $x=0$ seems to be a little higher than that for $%
x=-1$ at smaller $\delta $, but vice verse at larger $\delta $.
The critical density and the inflection pressure are larger for $x=0$ than $%
x=-1$ at low and moderate isospin asymmetries, and vice verse at larger $%
\delta $.

\subsection{Chemical instability}

Now we turn to discuss the chemical instability of hot asymmetric nuclear
matter. The chemical stability condition is
\begin{equation}
\left( \frac{\partial {\mu }_{n}}{\partial {\delta }}\right) _{P,T}>0\text{
and }\left( \frac{\partial {\mu }_{p}}{\partial {\delta }}\right) _{P,T}<0,
\label{Cstability}
\end{equation}%
and the system becomes chemical instable if either of the
inequalities in Eq. (\ref{Cstability}) is violated. For example,
if the isospin asymmetry $\delta $ has a small growth in the
region of chemical instability, it will even grow since more
neutrons will move into the system from the background, because
the low neutron chemical potential will lower the total energy of
the whole system. This also holds true for the case of protons. So
any isospin fluctuations will make the system unstable if  either
of the inequalities in Eq. (\ref{Cstability}) is not satisfied.
\begin{figure}[tbh]
\includegraphics[scale=0.8]{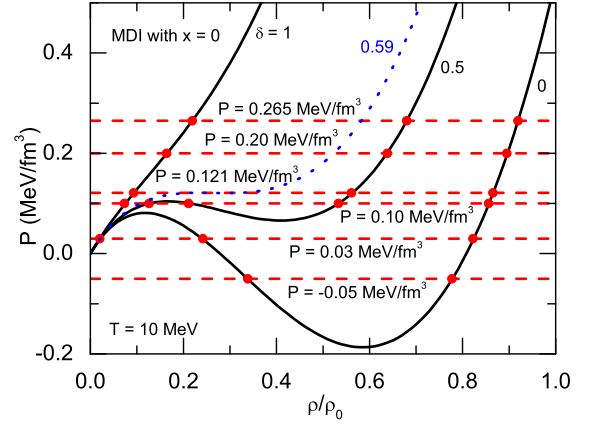}
\caption{{\protect\small (Color online) Pressure as a function of density at
fixed isospin asymmetry in the MDI interaction with $x=0$ at $T=10$ MeV to
show how to get the chemical potential isobar (see text for more details).}}
\label{example_DS1}
\end{figure}

To analyze the chemical instability, we calculate the chemical potential
isobar for neutrons and protons at fixed temperature and pressure. We do
this by searching for the cross point between the fixed pressure line and
the $P\sim \rho $ curves with fixed isospin asymmetry. In Fig.~\ref%
{example_DS1} we can calculate the densities and the chemical
potentials of the cross points, just for one isospin asymmetry at
fixed pressure and temperature. By changing the isospin asymmetry
from $0$ to $1$ thus get the whole chemical potential isobar at a
fixed temperature and pressure. We note that for different regions
of pressure, the number of the cross points can be one, two or
three for a fixed isospin asymmetry, which will be reflected in
the shape of the resulting chemical potential isobar. The critical
isospin asymmetry of mechanical instability is $0.59$ for the MDI
interaction at $T=10$ MeV with $x=0$ and the corresponding curve
in $P\sim \rho $ plane is plotted by dotted line in
Fig.~\ref{example_DS1}. The pressure of the inflection point is
$0.121$ MeV, above which the mechanical instability disappears and
only exists the chemical instability, and the chemical potential
isobar can only be one branch for all values of $\delta $.
\begin{figure}[tbh]
\includegraphics[scale=0.8]{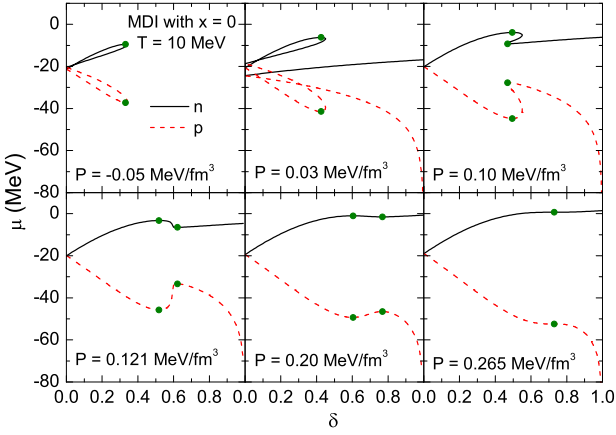}
\caption{{\protect\small (Color online) Chemical isobar as a function of the
isospin asymmetry in the MDI interaction with $x=0$ at $T=10$ MeV. The
extrema of each curve are indicated.}}
\label{example_DS2}
\end{figure}

Fig.~\ref{example_DS2} displays the chemical potential isobar
calculated at the pressure of $P=-0.05$, $0.03$, $0.10$, $0.121$,
$0.20$ and $0.265$ MeV$/$fm$^{3}$. One sees that the shape of the
curves is different for different pressures. The extrema of curves
are just the boundary of the chemical instability, or diffusive
spinodals (DS), which is indicated in Fig.~\ref{example_DS2} by
solid circles. $P=0.265$ MeV$/$fm$^{3}$ is the critical pressure
$P_{c}$ in the case of MDI interaction at $T=10$ MeV with $x=0$,
above which the chemical potential of neutrons (protons) increases
(decreases) monotonically with $\delta $ and the chemical
instability disappears. The inflection point which satisfies
\begin{equation}
\left( \frac{\partial {\mu }}{\partial {\delta }}\right) _{P_{c},T}=\left(
\frac{\partial ^{2}{\mu }}{\partial {\delta }^{2}}\right) _{P_{c},T}=0
\end{equation}%
is also plotted in the figure. We note here that the extrema of
$\mu _{n}$ and $\mu _{p}$ correspond to the same $\delta $ value
for the MDI and MID interaction and thus the critical pressure is
achieved simultaneously for neutrons and protons, but for the
eMDYI interaction the chemical potential isobar shows an
asynchronous behavior for neutrons and protons, as will be shown
in the following. This asynchronous behavior is also different for
different temperatures and values of $x$~\cite{Xu07b}.

The diffusive spinodals for the MDI, MID and eMDYI interactions
with $x=0$ and $x=-1$ at $T=5,10$ and $15 $ MeV are also shown in
Fig.~\ref{rhodelta1} and \ref{rhodelta2} in the $\rho\sim \delta $
plane, and in Figs.~\ref{Prho_T5}, \ref{Prho_T10} and
\ref{Prho_T15} in the $P\sim \rho $ plane. It is clearly shown
that the diffusive spinodal envelops the region of mechanical
instability and extends further out into the plane and the area of
chemical instability region decreases with increasing temperature.
It should be noted that the region of chemical instability is
between the curves of ITS and DS. At $\delta =0$, the DS and ITS
overlap for the MDI and eMDYI interaction, and the curves with
$x=0$ and $x=-1$ overlap as well. The boundary of chemical
instability is seen to be sensitive to the density dependence of
the symmetry energy. At $T=5$ MeV the maximum isospin asymmetry is
larger for $x=-1$ than $x=0$, while at $T=10$ and $15$ MeV it is
smaller for $x=-1$ than $x=0$. Furthermore, it is seen from
Figs.~\ref{rhodelta1} and \ref{rhodelta2} that the critical
isospin asymmetry and the area of chemical instability are also
sensitive to the isospin and momentum dependence of the nuclear
interaction, especially at higher temperatures. The maximum
$\delta $ value in the MDI interaction is similar to that in the
MID interaction at low temperature, but it becomes similar to that
in the eMDYI interaction at high temperature. This feature also
holds in the case of mechanical instability as discussed before.
In addition, the shape of the DS curve in the eMDYI interaction
with $x=-1$ at $T=5$ MeV observed in Figs.~\ref{rhodelta1},
\ref{rhodelta2} and \ref{Prho_T5} exhibits some unusual behaviors
and this is due to the asynchronous behavior of the chemical
potential isobar for neutrons and protons.

\section{Liquid-gas phase transition}

\label{transition}

\subsection{Chemical potential isobar}

\label{mudelta}

With the above theoretical models, we can now study the LG phase transition
in hot asymmetric nuclear matter. The phase coexistence is governed by the
Gibbs conditions and for the asymmetric nuclear matter two-phase coexistence
equations are
\begin{eqnarray}
P^{L}(T,\rho ^{L},\delta ^{L}) &=&P^{G}(T,\rho ^{G},\delta ^{G}),
\label{coexistenceP} \\
\mu _{n}^{L}(T,\rho ^{L},\delta ^{L}) &=&\mu _{n}^{G}(T,\rho ^{G},\delta
^{G}),  \label{coexistencemuN} \\
\mu _{p}^{L}(T,\rho ^{L},\delta ^{L}) &=&\mu _{p}^{G}(T,\rho ^{G},\delta
^{G}),  \label{coexistencemuP}
\end{eqnarray}%
where $L$ and $G$ stand for liquid phase and gas phase, respectively. The
Gibbs conditions (\ref{coexistenceP}), (\ref{coexistencemuN}) and (\ref%
{coexistencemuP}) for phase equilibrium require equal pressures and chemical
potentials for two phases with different concentrations and isospin
asymmetries. For a fixed pressure, the two solutions thus form the edges of
a rectangle in the proton and neutron chemical potential isobars as a
function of isospin asymmetry $\delta $ and can be found by means of the
geometrical construction method \cite{Mul95,Su00,Xu07b}.

\begin{figure}[tbh]
\includegraphics[scale=0.9]{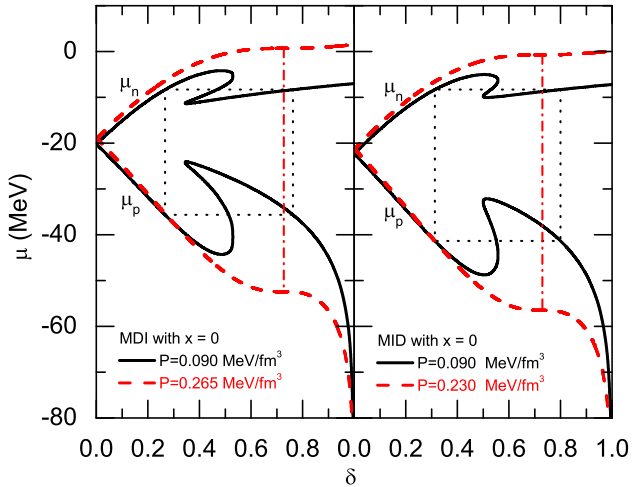}
\caption{{\protect\small (Color online) The chemical potential isobar as a
function of the isospin asymmetry $\protect\delta $ at $T=10$ MeV in the MDI
and MID interactions with $x=0$. The geometrical construction used to obtain
the isospin asymmetries and chemical potentials in the two coexisting phases
is also shown.}}
\label{mudeltamdimid}
\end{figure}

The way to calculate the chemical potential isobar is already explained in
above, and here we take the case of $T=10$ MeV as an example to show how we
study the LG phase transition from the chemical potential isobar. The solid
curves shown in Fig.~\ref{mudeltamdimid} are the proton and neutron chemical
potential isobars as a function of the isospin asymmetry $\delta $ at a
fixed temperature $T=10$ MeV and pressure $P=0.090$ MeV$/$fm$^{3}$ by using
the MDI and MID interactions with $x=0$. The resulting rectangles from the
geometrical construction are also shown by dotted lines in Fig.~\ref%
{mudeltamdimid}, from which one can see that different interactions give
different shapes for the chemical potential isobar. When the pressure
increases and approaches the critical pressure $P_{\mathrm{C}}$, an
inflection point will appear for proton and neutron chemical potential
isobars, i.e.,
\begin{equation}
\left( \frac{\partial {\mu }}{\partial {\delta }}\right) _{P_{\mathrm{C}%
},T}=\left( \frac{\partial ^{2}{\mu }}{\partial {\delta }^{2}}\right) _{P_{%
\mathrm{C}},T}=0.
\end{equation}%
Above the critical pressure, the chemical potential of neutrons (protons)
increases (decreases) monotonically with $\delta $ and the chemical
instability disappears. In Fig.~\ref{mudeltamdimid}, we also show the
chemical potential isobar at the critical pressure by the dashed curves. At
the critical pressure, the rectangle is degenerated to a line vertical to
the $\delta $ axis as shown by dash-dotted lines in Fig.~\ref{mudeltamdimid}
. The values of the critical pressure are $0.265$ and $0.230$ MeV$/$fm$^{3}$%
, respectively, for the MDI and MID interaction with $x=0$. It is
interesting to see that the different interactions give different values of
the critical pressure. We note that critical pressure is also sensitive to
the density dependence of nuclear symmetry energy with the stiffer symmetry
energy ($x=-1$) gives a smaller critical pressure \cite{Xu07b}.
\begin{figure}[tbh]
\includegraphics[scale=0.9]{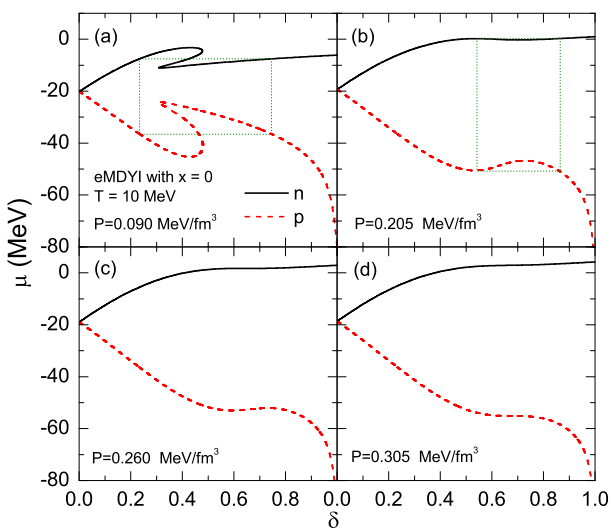}
\caption{{\protect\small (Color online) The chemical potential isobar as a
function of isospin asymmetry $\protect\delta $ at $T=10$ MeV in the eMDYI
interaction with $x=0$.}}
\label{mudeltamdyi}
\end{figure}

With the eMDYI interaction, the resulting single nucleon potential is
momentum dependent but its momentum dependence is isospin-independent.
Comparing the results with those of the MDI interaction, we can extract
information about the effects of the momentum dependence of the symmetry
potential while the effects of the momentum dependence of the isoscalar part
of the single nucleon potential can be studied by comparing the results with
those of the MID interaction. Shown in Fig.~\ref{mudeltamdyi} is the
chemical potential isobar as a function of the isospin asymmetry $\delta $
at $T=10$ MeV by using the eMDYI interaction with $x=0$. Compared with the
results from the MDI and MID interactions, the main difference is that the
left (and right) extrema of $\mu _{n}$ and $\mu _{p}$ do not correspond to
the same $\delta $ but they do for the MDI and MID interactions as shown in
Fig.~\ref{mudeltamdimid}. In particular, for the eMDYI interaction, the
chemical potentials of protons and neutrons are seen to exhibit asynchronous
variation with pressure. We note that this behavior depends on the
temperature and the value of $x$. At $T=10$ MeV the chemical potential of
neutrons increases more rapidly with pressure than that of protons. At $T=5$
MeV the chemical potential of neutrons increases more rapidly with $x=0$ but
vice verse with $x=-1$, while at $T=15$ MeV the asynchronous behavior seems
not quite obvious. This asynchronous variation is uniquely determined by the
special momentum dependence in the eMDYI interaction within the present
self-consistent thermal model. Actually, it is this asynchronous variation
that leads to the fact that the left (and right) extrema of $\mu _{n}$ and $%
\mu _{p}$ correspond to different values of $\delta $.

At lower pressures, for example, $P=0.090$ MeV/fm$^{3}$ as shown in Fig.~\ref%
{mudeltamdyi} (a), the rectangle can be accurately constructed and thus the
Gibbs conditions (\ref{coexistenceP}), (\ref{coexistencemuN}) and (\ref%
{coexistencemuP}) have two solutions. Due to the asynchronous variation of $%
\mu _{n}$ and $\mu _{p}$ with pressure, we will get a limiting pressure $%
P_{\lim }$ above which no rectangle can be constructed and the coexistence
equations (\ref{coexistenceP}), (\ref{coexistencemuN}) and (\ref%
{coexistencemuP}) have no solution. Fig.~\ref{mudeltamdyi} (b) shows the
case at the limiting pressure with $P_{\lim }=0.205$ MeV/fm$^{3}$ for $x=0$.
In this limit case, we note that the left edge of the rectangle actually
corresponds to the left extremum of $\mu _{p}$. With increasing pressure,
namely, at $P=0.260$, $\mu _{n}$ passes through an inflection point while $%
\mu _{p} $ still has a chemically unstable region and this case is shown in
Fig.~\ref{mudeltamdyi} (c). When the pressure is further increased to $%
P=0.305$ MeV/fm$^{3}$, as shown in Fig.~\ref{mudeltamdyi} (d), $\mu _{p}$
passes through an inflection point while $\mu _{n}$ increases monotonically
with $\delta $. We note that the asynchronous variation of $\mu _{n}$ and $%
\mu _{p} $ with pressure also depends on the value of $x$ \cite{Xu07b}.

\subsection{Binodal surface}

\begin{figure}[tbh]
\includegraphics[scale=0.8]{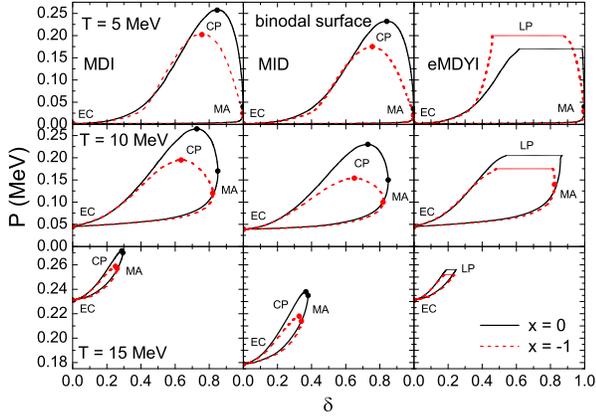}
\caption{{\protect\small (Color online) The binodal surface at $T=5$, $10$
and $15$ MeV in the MDI, MID and eMDYI interactions with $x=0$ and $x=-1$.
The critical pressure (CP), the limiting pressure (LP), the points of equal
concentration (EC) and maximal asymmetry (MA) are also indicated.}}
\label{Pdelta}
\end{figure}

For each interaction, the two different values of $\delta $ correspond to
two different phases with different densities and the lower density phase
(with larger $\delta $ value) defines a gas phase while the higher density
phase (with smaller $\delta $ value) defines a liquid phase. Collecting all
such pairs of $\delta (T,P)$ and $\delta ^{\prime }(T,P)$ thus forms the
binodal surface. Fig.~\ref{Pdelta} displays the binodal surface for the MDI,
MID and eMDYI interactions at $T=5$, $10$ and $15$ MeV with $x=0$ and $x=-1$
in the $P\sim \delta $ plain. As expected, for MDI and MID interactions the
binodal surface has a critical pressure, while for the eMDYI interaction the
binodal surface is cut off by a limit pressure. Above the critical pressure
or below the pressure of equal concentration (EC) point no phase-coexistence
region can exist. The EC point indicates the special case that symmetric
nuclear matter with equal density coexists, which is called ``indifferent
equilibrium" \cite{Mul95}. The maximal asymmetry (MA) also plays an
important role in LG phase transition. The left side of the binodal surface
is the region of liquid phase and the right side the region of gas phase,
and within the surface is the phase-coexistence region.

Now let's see what affects the critical or limit pressure and the
region of phase-coexistence. The critical pressure is sensitive to
the stiffness of the symmetry energy, with a softer symmetry
energy (with $x=0$) gives a higher critical pressure and a larger
area of phase-coexistence region. For the case of limit pressure
with eMDYI interaction this holds true at $T=10$ MeV and $T=15$
MeV, but the opposite result is obtained at $T=5$ MeV. For
symmetric nuclear matter the different values of $x$ give the same
EC point. As the MDI and eMDYI interactions are the same for
symmetric nuclear matter, they have the same EC point, but for the
MID interaction the EC point is lower, and the amount seems to
increase with increasing temperature. Below the limit pressure the
binodal surface is quite similar for the MDI and eMDYI
interactions. Comparing the MDI and MID interactions, the isospin
and momentum dependence seems to increase the critical pressure in
a larger amount. At $T=5$ MeV and $T=10$ MeV, the area of
phase-coexistence region for the MDI interaction is larger than
that for the MID interaction, but at $T=15$ MeV the opposite
result is observed. The critical or limit pressure seems not to
change monotonically with temperature, but it is clear that the
area of phase-coexistence region decreases with increasing
temperature and the pressure of EC point increases with increasing
temperature.

\begin{figure}[tbh]
\includegraphics[scale=0.8]{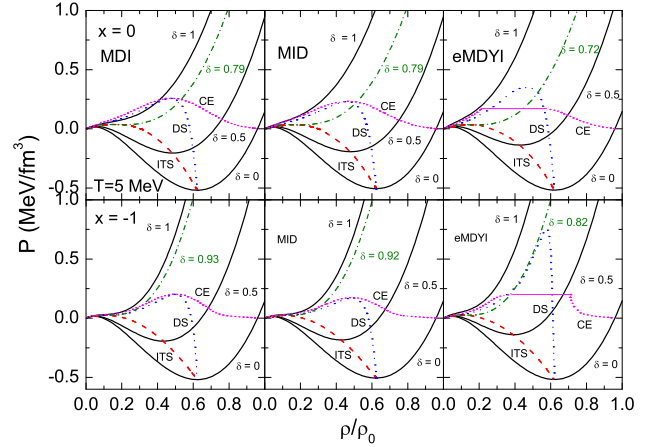}
\caption{{\protect\small (Color online) The phase coexistence boundary (CE)
in $P\sim\protect\rho$ plane for the MDI, MID and eMDYI interactions with $%
x=0$ and $x=-1$ at $T=5$ MeV. The isothermal spinodals (ITS) and diffusive
spinodals (DS) are also shown for comparison.}}
\label{PrhoT5}
\end{figure}

\begin{figure}[tbh]
\includegraphics[scale=0.8]{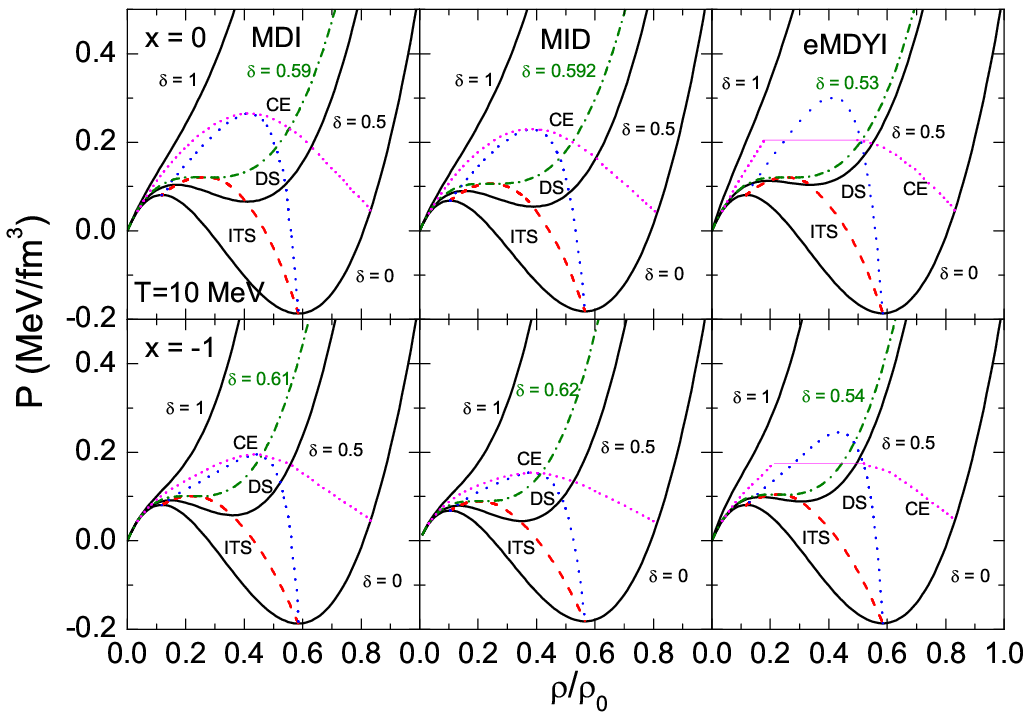}
\caption{{\protect\small (Color online) Same as Fig.~\protect\ref{PrhoT5}
but for $T=10$ MeV.}}
\label{PrhoT10}
\end{figure}

\begin{figure}[tbh]
\includegraphics[scale=0.8]{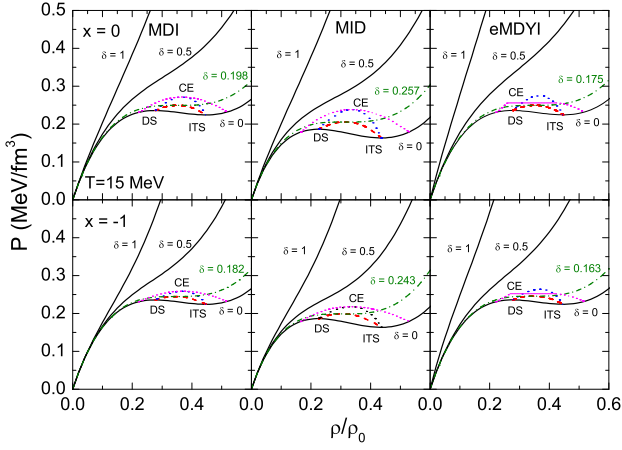}
\caption{{\protect\small (Color online) Same as Fig.~\protect\ref{PrhoT5}
but for $T=15$.}}
\label{PrhoT15}
\end{figure}

Fig.~\ref{PrhoT5}, \ref{PrhoT10}, \ref{PrhoT15} display the same
curves (the phase coexistence boundary (CE)) in $P\sim \rho $
plain, and the isothermal spinodals (ITS) and diffusive spinodals
(DS) are also included for comparison. For MDI and MID
interactions the critical pressure is the same for chemical
instability and binodal surface. For the eMDYI interaction as the
binodal surface is cut off by the limit pressure, the
phase-coexistence region can not extend the region of chemical
instability.

\subsection{Maxwell construction}

The binodal surface we show in the previous section provide much
information about the LG phase transition. As discussed in Ref.
\cite{Mul95}, we can analyze the process of LG phase transition in
hot asymmetric nuclear matter by Maxwell construction. We take the
case of MDI interaction with $x=-1$ at $T=10$ MeV as an example.

\begin{figure}[tbh]
\includegraphics[scale=0.8]{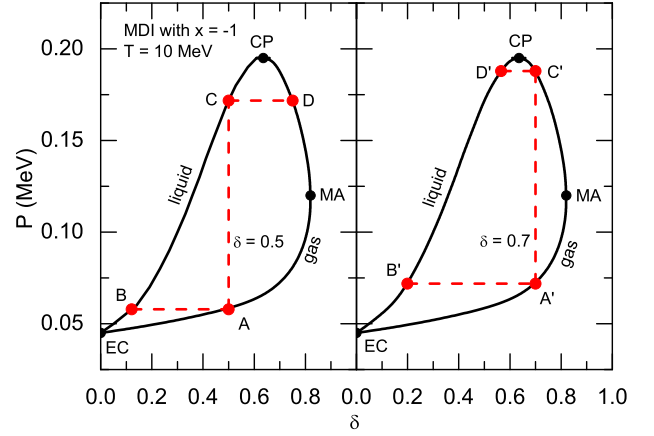}
\caption{{\protect\small (Color online) The binodal surface at $10$ MeV in
the MDI interactions with $x=-1$. The points A (A$^{\prime }$) through D (D$%
^{\prime }$) denote phases participating in a phase transition.
The critical pressure (CP), the points of equal concentration (EC)
and maximal asymmetry (MA) are also indicated.}} \label{PdeltaMDI}
\end{figure}

In the left panel of Fig.~\ref{PdeltaMDI} the system is compressed at a
fixed total isospin asymmetry $\delta =0.5$. The system begins from gas
phase, and encountered the two-phase region at the point A. Then a liquid
phase with higher density emerges from the point B, with infinitesimal
proportion. As the system is compressed, the gas phase evolves from A to D,
while the liquid phase evolves from B to C. In this process the gas phase
and the liquid phase coexist and the proportion of each phase changes, but
the total isospin asymmetry is fixed. At the point C the system totally
changes from gas phase to liquid phase and leaves the phase-coexistence
region.

\begin{figure}[tbh]
\includegraphics[scale=0.8]{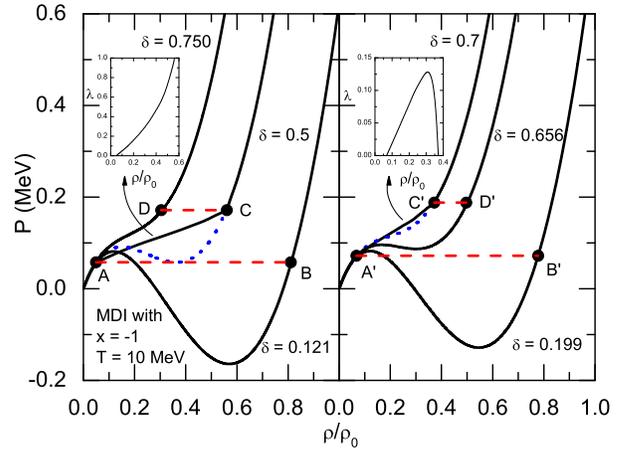}
\caption{{\protect\small (Color online) The LG phase transition process in
the $P\sim \protect\rho $ plain for the MDI interaction with $x=-1$ at $T=10$
MeV. The system is initially compressed at fixed total isospin asymmetry $%
\protect\delta =0.5$ (left panel) and $\protect\delta =0.7$ (right panel).
The Maxwell construction produces the curve AC (A$^{\prime }$C$^{\prime }$).
The inset displays the fraction of the liquid phase $\protect\lambda $ from
A(A$^{\prime }$) to C(C$^{\prime }$).}}
\label{PrhoMDI}
\end{figure}

We can analyze this process in the phase-coexistence region by solving the
following equations
\begin{eqnarray}
\lambda \delta ^{L}\rho ^{L}+(1-\lambda )\delta ^{G}\rho ^{G} &=&\delta \rho,
\label{maxwell1} \\
\lambda \rho ^{L}+(1-\lambda )\rho ^{G} &=&\rho,  \label{maxwell2}
\end{eqnarray}%
where $\delta ^{L(G)}$ and $\rho ^{L(G)}$ are the isospin
asymmetry and density of liquid (gas) phase. The total isospin
asymmetry $\delta $ in this case is $0.5$. The fraction of the
liquid phase $\lambda $ and the total density $\rho $, from which
Maxwell construction can be produced, can be obtained by solving
the above equations. The corresponding isotherms are drawn in the
left panel of Fig.~\ref{PrhoMDI}. The dotted line connecting A and
C obtained by direct calculation is unphysical. The nearly
straight line connecting A and C is produced by Maxwell
construction and corresponds to the realistic process. The
fraction of the liquid phase $\lambda $ from A to C is also shown
in the inset and it changes monotonically from $0$ to $1$.

The geometry of the binodal surface offers a second possibility for the LG
phase transition process. The situation is displayed in the right panel of
Fig.~\ref{PdeltaMDI}, where the system is compressed at fixed total isospin
asymmetry $\delta =0.7$, which is larger than the isospin asymmetry of the
CP point. The system begins from gas phase, and encountered the two-phase
region at the point A$^{\prime }$. Then a liquid phase with infinitesimal
fraction emerges from the point B$^{\prime }$. As the system is compressed,
the gas phase evolves from A$^{\prime }$ to C$^{\prime }$, while the liquid
phase evolves from B$^{\prime }$ to D$^{\prime }$. The system crosses the
phase-coexistence region, but at the point C$^{\prime }$ it remains at gas
phase and leaves the binodal surface on the same branch. The corresponding
isotherms are shown in the right panel of Fig.~\ref{PrhoMDI}. The solid line
rather than the dotted one connecting A$^{\prime }$ and C$^{\prime }$
corresponds to the real process of LG phase transition. In this case the
fraction of the liquid phase $\lambda $ increases from $0$ to $\lambda _{max}
$ (about $0.13$) and then drops to $0$ again.

\subsection{Order of LG phase transition}

\begin{figure}[tbh]
\includegraphics[scale=0.8]{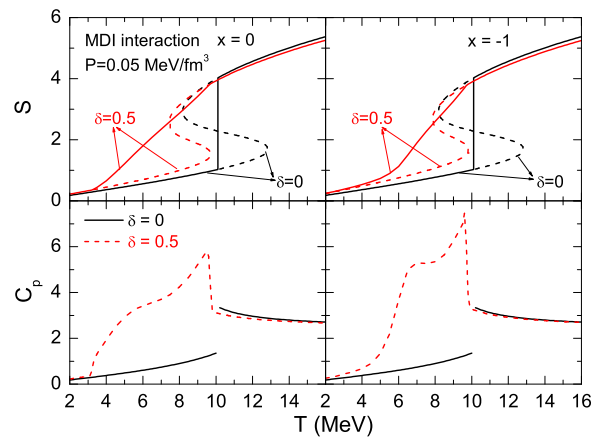}
\caption{{\protect\small (Color online) Evolution of entropy per nucleon and
specific heat per nucleon with temperature under fixed pressure $0.05$MeV$/$%
fm$^{3}$ at $\protect\delta =0$ and $0.5$ for the MDI interaction
with $x=0$ (left panels) and $x=-1$ (right panels). In the upper
panels, the dashed line is obtained by direct calculation while
the solid line is from Maxwell construction.}}
\label{SCpTMDIP005}
\end{figure}

In the following we consider the order of LG phase transition and focus on
the realistic MDI interaction by observing the behavior of thermodynamical
quantities under fixed pressure. Here we use the pressure of $P=0.05$ MeV$/$%
fm$^{3}$ and we note that there are no qualitative changes if
other pressures below the critical pressure are used. We choose a
relatively low pressure just to see more clearly the effects of
the phase transition on the thermodynamical quantities. In the
upper panels in Fig.~\ref{SCpTMDIP005} we show the evolution of
entropy per nucleon with temperature at fixed pressure $0.05$
MeV$/$fm$^{3}$ for isospin asymmetries of $\delta =0$ and $0.5$
using the MDI interaction with $x=0$ and $x=-1$ respectively. The
method is quite similar to calculating the chemical potential
isobar above, so we do not go into the details about the
calculation method. In the upper panels in Fig.~\ref{SCpTMDIP005},
the dashed line is obtained by direct calculation and it is
unphysical, while the solid line corresponds to the realistic
process and is obtained by Maxwell construction. We calculate the
chemical potential isobar at every temperature of the
phase-coexistence region under fixed pressure, then find the
densities and isospin asymmetries of the coexistence phase from
Gibbs conditions, and thus obtain the fraction of each phase by
using Eqs. (\ref{maxwell1}) and (\ref{maxwell2}). The total
entropy per nucleon $S $ in the coexistence phase is then
calculated from
\begin{equation}
S(\rho ,\delta ,T)=\lambda S^{L}(\rho ^{L},\delta ^{L},T)+(1-\lambda
)S^{G}(\rho ^{G},\delta ^{G},T),  \label{entropy}
\end{equation}%
where $S^{L(G)}$ can be obtained from $\rho ^{L(G)}$ and $\delta
^{L(G)}$ by using Eq.~(\ref{S}). From the upper panels in
Fig.~\ref{SCpTMDIP005} we can see that at $\delta =0$ the entropy
jumps at $T=10.1$ MeV, which clearly indicates that the LG phase
transition for symmetric nuclear matter under the pressure of $0.05$MeV$/$fm$%
^{3}$ (which is below the critical pressure) is of first order.
The transition temperature is $T_{c}=10.1$ MeV, and we note that
this value depends on the fixed pressure used. On the other hand,
the curves of $\delta =0.5$ is continuous. To be more specific, we
calculate the heat capacity per nucleon under fixed pressure from
\begin{equation}
C_{p}(T)=T\left( \frac{\partial S}{\partial T}\right) _{P,\delta
},\label{heat capacity}
\end{equation}%
where the entropy per nucleon is obtained from Eq.
(\ref{entropy}). The lower panels in Fig.~\ref{SCpTMDIP005}
display the heat capacity per nucleon as a function of temperature
under fixed pressure $0.05$ MeV$/$fm$^{3}$ for isospin asymmetries
of $\delta =0$ and $0.5$ using the MDI interaction with $x=0$ and
$x=-1$ respectively. For both cases of $x=0$ and $x=-1$ at $\delta
=0.5$, the heat capacity is continuous but its first derivative is
not continuous, which indicates that the LG phase transition for
asymmetric nuclear matter is of second order according to
Ehrenfest's definition of phase transitions \cite{Rei80}. These
results are consistent with those in Ref. \cite{Mul95} with a
different model. Although here we only consider the MDI
interaction, we note that the order of LG phase transition does
not depend on the isospin and momentum dependence of the nuclear
interaction.

\section{Summary}

\label{summary}

Within a self-consistent thermal model we have studied in detail
and systematically effects of the isospin and momentum dependent
interactions on thermal properties of asymmetric nuclear matter.
We put the special emphasis on the temperature dependence of the
isospin-dependent bulk and single-particle properties, the
mechanical and chemical instabilities, and the LG phase transition
of the hot asymmetric nuclear matter. In our analyses we used the
isospin and momentum dependent MDI interaction constrained by the
isospin diffusion data in heavy-ion collisions, the
momentum-independent MID interaction, and an isoscalar
momentum-dependent eMDYI interaction. Our results indicate that
the EOS and the symmetry energy are quite similar for the three
interactions at finite temperature, which implies that their
temperature dependence is not so sensitive to the momentum
dependence of the interaction. In particular, the symmetry energy
at fixed density is found to generally decrease with temperature
for the three interactions. For the MDI interaction, both the
kinetic and potential parts of the symmetry energy are temperature
dependent and the decrement of the symmetry energy with
temperature is essentially due to the decrement of the potential
energy part of the symmetry energy with temperature. On the other
hand, for the MID and eMDYI interactions, the decrement of the
symmetry energy with temperature is only due to the kinetic part
of the symmetry energy since the potential part does not depend on
the temperature. Compared with the MID and eMDYI interactions, the
single-particle potential of the MDI interaction is isospin and
momentum dependent, which leads to the fact that the symmetry
potential is momentum dependent and the nucleon effective mass is
splitted in asymmetric nuclear matter for the MDI interaction. It
is further shown that only the low momentum part of the
single-particle potential and the nucleon effective mass is
significantly lifted with increasing temperature for the MDI and
eMDYI models. For the MDI interaction, the low momentum part of
the symmetry potential is significantly reduced with increasing
temperature.

We have also analyzed the boundaries of both mechanical and
chemical instabilities. We found that the area of both mechanical
and chemical instabilities generally decreases with increasing
temperature. Meanwhile, the boundaries are shown to be sensitive
to the density dependence of nuclear symmetry energy. In the case
of mechanical instability the critical isospin asymmetry is larger
for stiffer symmetry energies (e.g., $x=-1$) at low and moderate
temperatures and vice verse at high temperatures. While for the
chemical instability the maximum asymmetry is larger for stiffer
symmetry energies at low temperatures and vice verse at high and
moderate temperatures. Furthermore, it is indicated that the
mechanical and chemical instabilities are also sensitive to the
isospin and momentum dependence of the nuclear interaction,
especially at higher temperatures. The boundaries obtained with
the MDI interaction are similar to those obtained with the MID
interaction at low temperatures and with the eMDYI interaction at
high temperatures.

Finally, we have explored in detail the effects of isospin and
momentum dependence of the nuclear interaction on the liquid-gas
phase transition in hot asymmetric nuclear matter. The boundary of
the phase-coexistence region is shown to be sensitive to the
density dependence of the nuclear symmetry energy with a softer
symmetry energy giving a higher critical pressure and a larger
area of phase-coexistence region. The critical pressure and the
area of phase-coexistence region is also quite sensitive to the
isospin and momentum dependence of the nuclear interaction by
comparing the cases of MDI and MID interactions. For the eMDYI
interaction, a limiting pressure above which the LG phase
transition cannot take place has been found and it is shown to be
sensitive to the stiffness of the symmetry energy as well.
Furthermore, the area of phase-coexistence region decreases with
increasing temperature, and the pressure of the EC point increases
with increasing temperature for all of the three interactions
considered here. The phase transition process and the order of
phase transition have also been analyzed by using the Maxwell
construction. Our results indicate that the LG phase transition
for symmetric nuclear matter is of first order but it becomes
second order for the asymmetric nuclear matter.

\begin{acknowledgments}
This work was supported in part by the National Natural Science Foundation
of China under Grant Nos. 10334020, 10575071, and 10675082, MOE of China
under project NCET-05-0392, Shanghai Rising-Star Program under Grant No.
06QA14024, the SRF for ROCS, SEM of China, the China Major State Basic
Research Development Program under Contract No. 2007CB815004, the US
National Science Foundation under Grant No. PHY-0652548 and the Research
Corporation under Award No. 7123.
\end{acknowledgments}

\end{document}